\documentclass[preprint,12pt]{elsarticle}

\usepackage{amssymb}
 \usepackage{amsthm}
 
 \usepackage{amsmath}

\journal{Extreme Mechanics Letters}

\begin{document}

\begin{frontmatter}



\title{Liquid Structures: a novel Computational Fluid Dynamics (CFD)  inspired metamaterial}


\author[l1]{Paolo Gallina}
%%Lorenzo Scalera, Massimiliano Gei and Stefano Seriani
\address[l1]{University of Trieste, Department of Engineering and Architecture, Via Valerio 10, 34127 Trieste, Italy}
\author[l1]{Massimiliano Gei}
%\address[l2]{DICAM, University of Trento, Via Mesiano 77, 38123 Trento, Italy}
%\address[l3]{School of Engineering, Cardiff University, The Parade, Cardiff CF24 3AA, UK}
\author[l4]{Lorenzo Scalera}
\author[l1]{Stefano Seriani}
\address[l4]{University of Udine, Polytechnic Department of Engineering and Architecture, Via delle Scienze 206, 33100 Udine, Italy}

\begin{abstract}
%This paper presents new metamaterial structures we call %\textit{liquid structures}: a topology of bistable %mechanisms made up of a high number of sub-mechanisms, named \textit{cells}. Cells are composed of pseudo-rigid links and joints (prismatic or revolute). The name \textit{liquid structures} comes from the similarities they present with the constant flow of incompressible fluids they are inspired to. A general theoretical framework is presented as well as a detailed discussion about future theoretical and mechanical challenges arisen by this new metamaterial paradigm.

We present a theoretical framework for new structural metamaterials we refer to as \textit{liquid structures}: a topology of bistable mechanisms made up of a high number of \textit{cells} that are sub-mechanisms composed of pseudo-rigid links and joints. The name \textit{liquid structures} comes from the similarities they present with the kinematics of the constant flow of incompressible fluids they are inspired by in a limited domain. The layout of the cells are obtained through a two-step process where: (i) the node displacements are computed by means of a Computational Fluid Dynamics tool feeding and (ii) the kinematic synthesis of each cell that is subsequently performed. We report two illustrative examples where star- and diamond-type cells are employed. The paper concludes with a detailed discussion about future theoretical and manufacturing challenges arising from this new metamaterial paradigm.
\end{abstract}


%%\item Research highlight 1
%%\item Research highlight 2


\end{frontmatter}


\section{Introduction}
Metamaterials are materials engineered to show global properties (mechanical, electrical, optical, etc.) that are not found in naturally bulk materials \cite{Pendryschurig,Maldovan,LiGao,Kosal2020135}. They present unconventional behaviours, both for 2D and 3D applications \cite{Goswami2019, Bertoldiflex,reviewcinesi}. 
%A clear and updated overview about mechanical meta-materials can be found in \cite{Barchiesi2019212}.
Among these, metamaterials based on lattice structures have been deeply studied, especially for their mechanical properties (for example, negative Poisson's ratio and bistability \cite{Babaee20135044,Cabras16}), or for their performance in dynamics \cite{Chen2020,An2019,Hu2019,Garau19}. %Moreover, vibrational modes are related to surface phonons that present some similarities with electronic quantum matter \cite{Bansil2016}.
Their versatility has facilitated their use in other scientific fields such as biology (e.g., knotted proteins and DNA \cite{Lua2006350}, higher genus membranes \cite{Michalet1995666}) and physics (e.g. topological study of the behaviour of defects in liquid crystals \cite{Alexander2012497}).

Generally, lattice-based metamaterials can be seen as a network of nodes (with or without mass) connected by spring-like elastic elements that can act either as a strut (in compression) or a tie (in tension) \cite{Calladine78,Vitelli201212266}. Such elastic elements may also consist of pseudo-elastic links \cite{pseudolink} --a review on pseudo-elastic rigid bodies can be found in \cite{Howell2016}. It is intuitive that the overall rigidity of the structure is on average related to the number of nodes that are connected to each single node. On the one hand, reducing the number of links makes the structure less rigid, until a critical point is reached  at which the overall assembly can be infinitesimally deformed without any relevant energy cost, i.e. the structure loses rigidity. On the other, a not-carefully checked distribution of a sufficient number of nodes and links may still display internal Degrees of Freedom (shortened henceforth as \#DoF) or, as recently defined, \textit{soft modes}. 
%
%Although the network topology is key to foresee the existence of \textit{soft modes} (zero energy displacements), they depend sensitively on how the nodes are distributed spatially throughout \cite{Sun201212369}. 
%
In other words, the stability of a structure can not be evaluated by considering exclusively topological aspects. 
This is the reason why, as far as linkage mechanisms are concerned, the Grubler's equation that provides the number  \#DoF of a mechanism fails if exceptions are not taken into account \cite{Makkonen1994145}.

For example, with reference to Kagome lattices, Vitelli explained in a clear way how topology is not everything:
``The character of the soft modes depends sensitively on boundary conditions and network architecture (e.g., on the relative angle between bonds and not merely on the average coordination number)" \cite{Vitelli201212266}.
%At the heart of such kinematic behaviour is often a \textit{soft mode}: a motion that does not significantly stretch or compress the links between nodes.
This principle is also exploited in providing mobility in MEMS (Micro Electro-Mechanical Systems) since, at the microscopic scale, it is not possible to realize complex mechanical transmissions by gears or shafts. In some cases, it is feasible to build complex structures that, for small displacements, allow the synchronized motion of several bodies using a single actuator. For example, Gallina et al. have realized a platform for biological cell tests (multiaxial stretcher), actuated by an electrostatic comb \cite{Scuor2006239}. The upgrade of the system  was a multi-axis actuator consisting of an apparently redundant structure. In this case, the mobility was guaranteed by a special linkage that provided a soft mode 
\cite{Antoniolli2014131}.
Paulose et al. were able to create topological isolated soft modes positioned at desired locations in a metamaterial composed of either Kagome or square lattices. Their structures are robust against structural deformations \cite{Paulose2015153} and this relies on the concept of \textit{topological polarization}. A topologically polarized lattice can realize soft modes of self-stress at sample edges \cite{Kane201339}.

We propose here a different approach for the design and control of soft modes that combines techniques used in mechanisms synthesis (linkage synthesis) with
Computational Fluid Dynamics (CFD). 
These tools allow us to
link one boundary location of a lattice structure to a far away one through paths where lattice displacements are ``channeled" to ensure the required actuation.

%create soft modes that spread along narrow and connected ``channels" so as to kinematically . 

%These tools allow us to create soft modes that spread along narrow and connected ``channels" so as to kinematically link one boundary location of the structure to a far away one.  

\begin{figure}[h!]
\centering
    \includegraphics[width=0.6\textwidth]{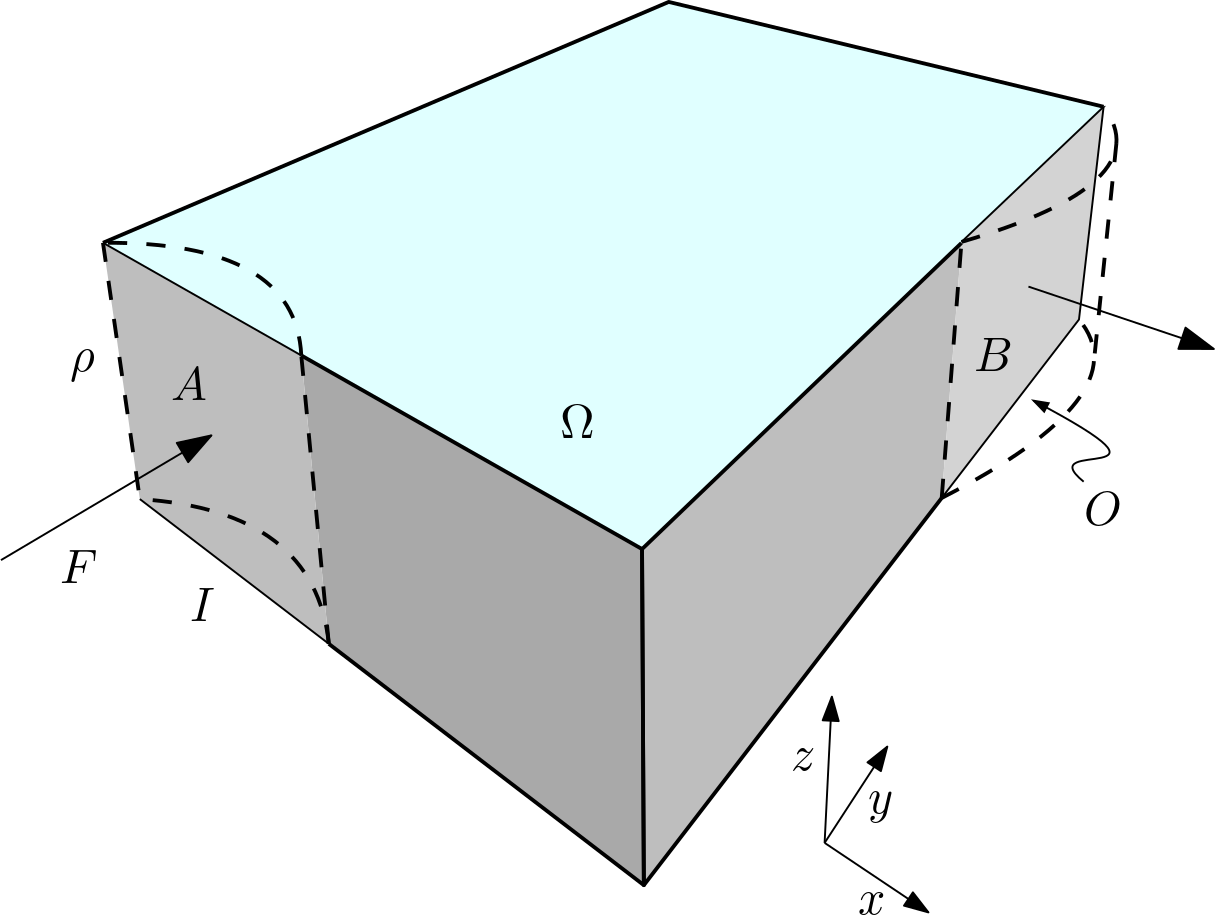}
    \caption{Sketch of the volume $\Omega\times \rho$ that is occupied by  the liquid structure. When $A=I\times \rho$ is pressed, the distant surface $B=O\times \rho$ is deformed.}
    \label{fig:parallelepippedo}
\end{figure}

Therefore, the paper is devoted to a new family of metamaterial linkages we refer to as \textit{liquid structures}. 
In order to describe the principle of operation, we introduce the following mechanical problem: consider a simply connected domain $\Omega$ in the plane $<x,y>$, bounded by a perimeter $p$, as in Fig. \ref{fig:parallelepippedo}. In the example, $\Omega$ is a rectangle and  $\Omega\times \rho$ a volume, where $\rho$ is a given thickness.
The parallelepiped is occupied by a complex indeterminate mechanical system that must exhibit the following property: if one portion of its lateral surface is deformed by applying a load, the system must react by deforming a distant surface. In Fig. \ref{fig:parallelepippedo} the load, equivalent to the force $F$, is applied to $A=I \times \rho$ ($I \subset p$) whereas the remote surface is 
$B = O \times \rho$ ($O \subset p$). To fulfill the objective, $A$ and $B$ must be related by a sort of mechanical link, so that a concave deformation of $A$ causes the convex deformation of $B$.
How can such a system be implemented?

One possibility to achieve the goal is to create a hydraulic press: an empty parallelepiped filled with incompressible fluid, provided with rigid walls except at surfaces $A$ and $B$; $A$ and $B$ are equipped with two deformable and sealed membranes to prevent the liquid from leaking. In this way, if membrane $A$ is pressed, due to the incompressibility of the fluid, the membrane $B$ deforms accordingly to maintain the deformation isochoric.
The question that inspired this research work is: \textit{is it possible to create a mechanism composed of compliant links and joints, as general as possible, able to behave like the liquid system, at least for small displacements}? 

To provide a positive answer, a new family of mechanisms, that of  \textit{liquid structures}, is conceived: a liquid structure is made up of links and  (revolute and prismatic) joints that connect the displacements of some points (on the face $A$ in Fig. \ref{fig:parallelepippedo}) to those of other points located at a different spot (on the face $B$).

We prefer the definition of \textit{liquid structure} instead of \textit{liquid mechanism} since this new linkage allows only small displacements and the majority of its points on the border are fixed to the reference frame. The parallelism between the behavior of a box filled with liquid and our linkage  explains the metaphorical term \textit{liquid} employed to address this new family of mechanisms.

To the best of our knowledge, this is the first study that proposes the synthesis of a lattice to create a ``mechanical transmission" within the network. It is also the first work that takes advantage of CFD techniques to perform the kinematic synthesis. The great availability of CFD codes allows to set in a simple way the input data of the problem and get an insight into a possible solution.

In the next section of the paper a general theoretical framework is given where star- and diamond-type cells are also introduced. Sections 3 and 4 show how to apply the method to a couple of case studies. In Section 5, open questions and future developments are presented. Eventually, concluding remarks are drawn in Section 6. 

\section{Mathematical framework} \label{Mathematical framework}
\subsection{Model framework and definitions}
\begin{figure}[h!]
\centering
    \includegraphics[width=0.95\textwidth]{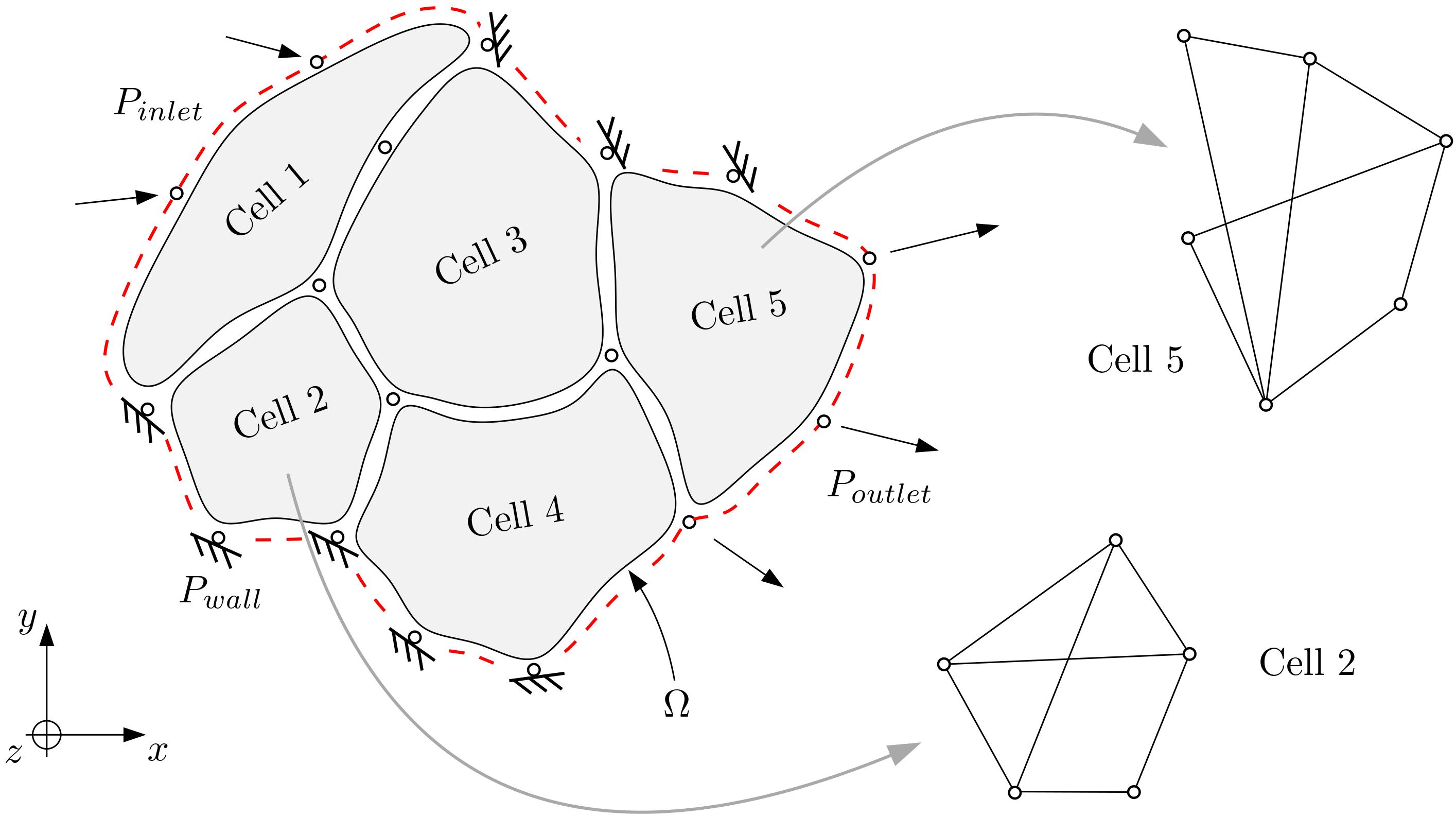}
    \caption{Schematic representation of the cell partition of a 2D liquid structure occupying the domain $\Omega$. $z$ is the  out-of-plane axis, pointing outward.  The internal arrangements of two representative cells are displayed on the right. Cell 2 is internally rigid whereas cell 5 possesses an internal DoF. }
    \label{fig:area}
\end{figure}

To illustrate the theoretical framework, we refer our developments to the two-dimensional case; therefore, the thickness $\rho$ is not taken into account. Consider the area \( \Omega \in \mathbb{R}^2 \) of Fig. \ref{fig:area} that can be  ``filled'' with a complex linkage mechanism made up of revolute and prismatic joints\footnote{\lq Hinges' and \lq sliders' can be used as an alternative of \lq revolute' and \lq prismatic' joints, respectively.} to  form a complex lattice that can be split into sub-mechanisms we refer to as \textit{cells}.

Each cell, regarded as not connected to the rest of the linkage, may have 3, 4 or even more than 4 DoF, and is connected to the adjacent ones through points that are called \textit{nodes}. Cells with 3 DoF are rigid and have null mobility: they can freely move in the plane as a rigid body (by translating along axes $x$ and $y$ and rotating about axis $z$), but do not deform internally.
Those with 4 DoF possess an internal degree of freedom in addition to the possibility of rigid motion.

The number of nodes for each cell is $n_j$\footnote{It is worth pointing out that the same node may also belong to an adjacent cell.}, where the subscript $j$ indicates the \textit{j}-th cell.   
We define four different kinds of nodes, the first three of which belong to the boundary of  \( \Omega\): 
\begin{itemize}
\item \textit{Wall nodes}: nodes that constitute the fixed part of the boundary of the domain which is in turn fixed to the inertial frame $<x,y>$. They are indicated with the symbol $P_{wall}$ and $n_{wall}$ is their number.
\item \textit{Inlet nodes}: nodes that can displace inwardly in \( \Omega\). They are indicated with the symbol $P_{inlet}$ and $n_{inlet}$ is their number.
\item \textit{Outlet nodes}: nodes that can displace outwardly in \( \Omega\). They are indicated with the symbol $P_{outlet}$. $n_{outlet}$ is the number of outlet nodes.
\item \textit{Internal nodes}: they are inner points belonging to \( \Omega\). They are indicated with the symbol $P_{in}$. $n_{in}$ is their number.
\end{itemize}

In Fig. \ref{fig:area}, just the boundaries (nodes) of the cells are represented, not their sub-mechanisms. As an example, the sub-mechanism of cell 5 is sketched at the right-hand side. It is made up of a combination of two 4-bar linkages and possesses 4 DoF, as it can be easily inferred by the Gruebler's equation:
\begin{equation} \label{grubler_1}
   n_{DoF_j}=3\, nl_{j}-2\, nc_{j}=3\times8-2\times10=4, 
\end{equation}
where $n_{DoF_j}, nl_{j},$ and $nc_{j}$ represent the \#DoF, the number of links and the number of nodes of the $j$-th cell, respectively. In this specific case, $j=5$.
As observed in Fig. \ref{fig:area}, cell 2 is internally rigid, since 
\begin{equation} 
   n_{DoF_2}=3\, nl_{2}-2\, nc_{2}=3\times7-2\times9=3. 
\end{equation}

The goal of the theory is to make the synthesis of each cell sub-mechanism to create a kinematic relationship between inlet and outlet nodes, as explained in the next sub-section.

The problem is defined as follows: given the displacements $\delta_{inlet\,j}$ ($j\in\{1,...,n_{inlet}\}$) imposed to the inlet nodes $P_{inlet\,j}$ and the displacements $\delta_{outlet\,j}$ ($j\in\{1,...,n_{outlet}\}$) imposed to the outlet nodes $P_{outlet\,j}$, calculate the displacements $\delta_{in\,j}$ ($j\in\{1,...,n_{in}\}$) of the internal nodes $P_{in\,j}$ and perform the kinematic synthesis of each cell.

\subsection{Kinematic synthesis procedure}
The procedure to perform the kinematic synthesis of the entire liquid structure is split into two steps as detailed in the next two sub-sections. In STEP 1 the node displacements are calculated, whereas STEP 2 provides the kinematic synthesis of each cell.

\subsubsection{STEP 1: node displacements calculation} \label{STEP_1}
Let us assume that the liquid structure has $n_P=n_{wall}+n_{inlet}+n_{outlet}+n_{in}$ nodes and $n_c$ cells.
Input of the STEP 1 problem is:
\begin{itemize}
\item the initial coordinates of each point $P_j$ of the domain, where $j\in\{1,...,n_P\}$;
\item the inlet node displacements $\delta_{inlet\,j}$ ($j\in\{1,...,n_{inlet}\}$);
\item the outlet displacements $\delta_{outlet\,j}$, where  $j\in\{1,...,n_{outlet}\}$.
\end{itemize}
The output of the problem is the calculation of the internal node displacements $\delta_{in\,j}$ ($j\in\{1,...,n_{in}\}$).

\begin{figure}[h!]
\centering
    \includegraphics[width=0.75\textwidth]{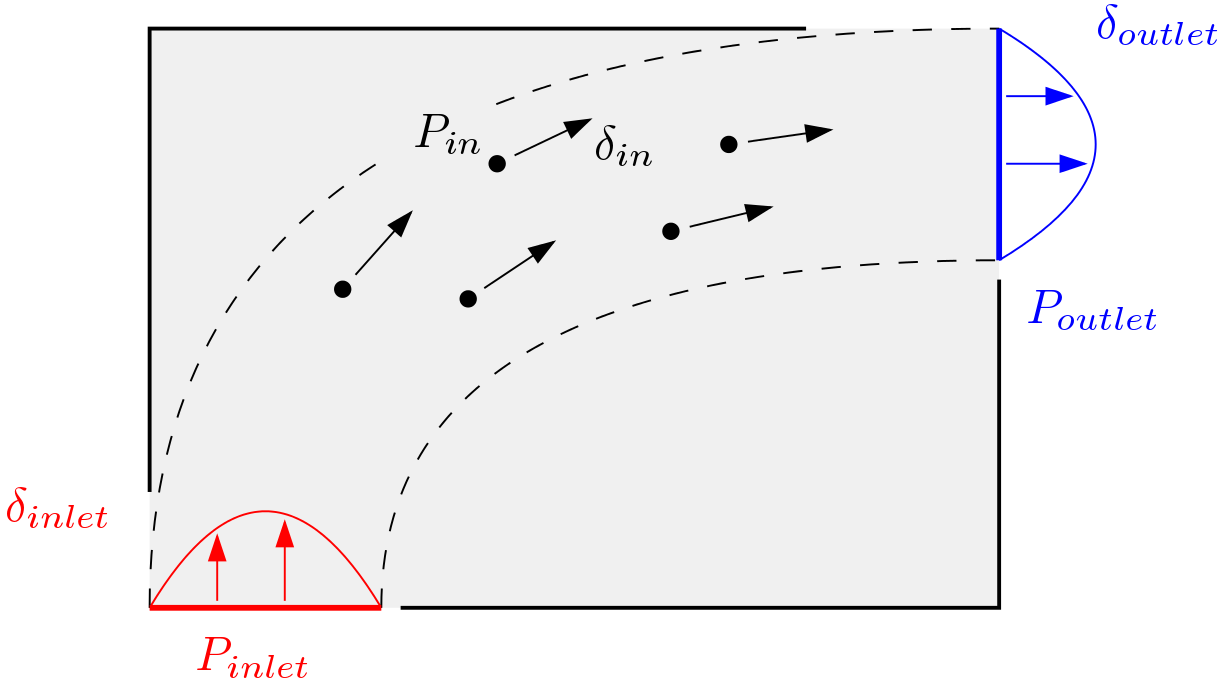}
    \caption{Sketch illustrating how the displacements of nodes within the liquid structure smoothly change along a ``corridor area" joining inlet and outlet segments.}
    \label{fig:corridor}
\end{figure}

To reach the goal, different strategies can be pursued. Speaking in qualitative terms, whatever the approach, it must be ensured that the displacements $\delta_{in\,j}$ progressively and smoothly change along a ``corridor area" that joins the inlet boundary to the outlet one (see Fig. \ref{fig:corridor}). What should be the shape of this corridor and which criterion to use to calculate the displacements remain an interesting and challenging open problem.

We solve the issue by exploiting a sort of parallelism between liquid structures and the dynamic behaviour of fluid flow inside the domain $\Omega$ characterized by an inlet and an outlet. In fact, it is possible to perform a CFD computation -- CFD resolves Navier–Stokes equations-- in order to derive the node displacements. More details will be provided in Section \ref{implementation}.

\begin{figure}[h!]
\centering
    \includegraphics[width=0.99\textwidth]{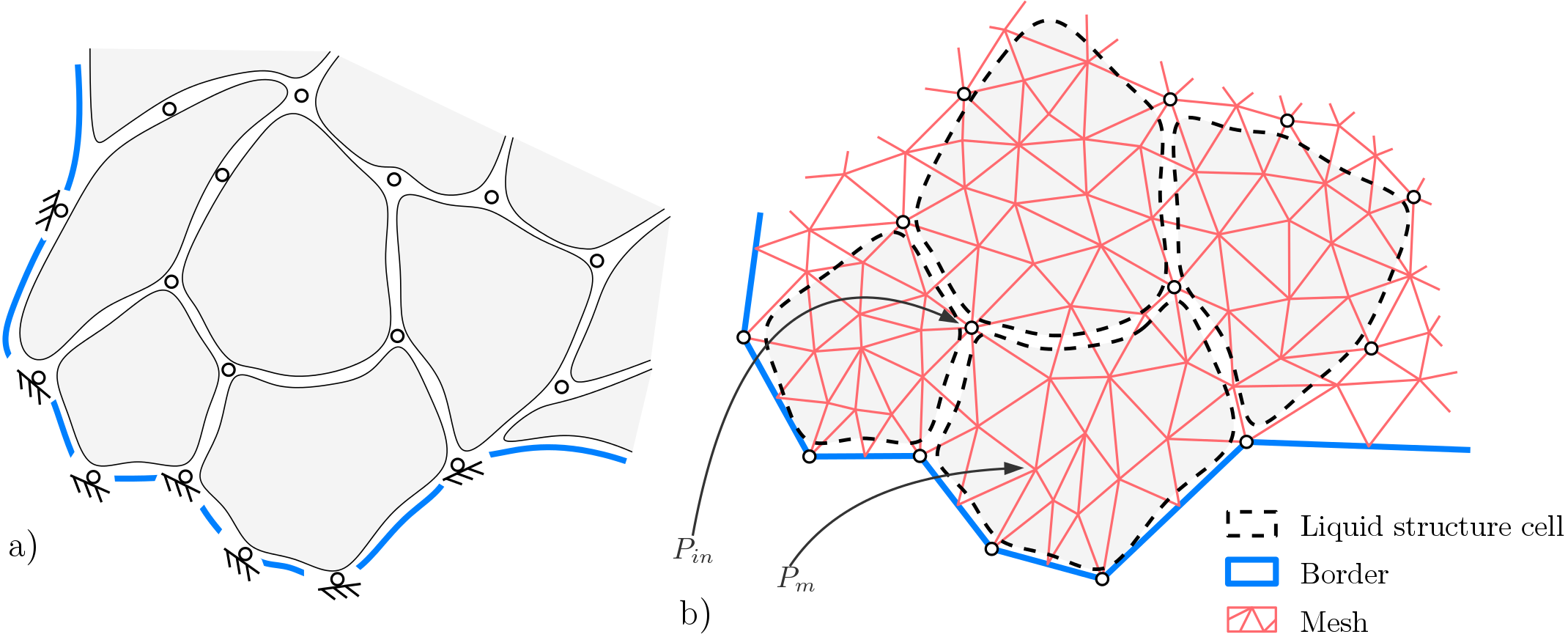}
    \caption{a) Sketch of the physical model of liquid structure; b) Mesh generated by the CFD computation tool that involves a larger number of nodes than that of the physical ones.}
    \label{fig:mesh}
\end{figure}

It is assumed that:
\begin{itemize}
\item the fluid is water (however, similar results can be obtained with other fluids); 
\item the flow regime is Steady State;
\item the process is isothermal;
\item the flow is laminar.
\end{itemize}
Still in the 2D framework, the domain of the CFD problem is $\Omega$, the inlet is given by segment $I$ and the outlet by segment $O$. Given the geometry, a mesh is generated (as illustrated in Fig. \ref{fig:mesh}). 
A generic node of the mesh is named $P_m$. Let us assume that the mesh nodes are distributed equispatially along the perimeter $p$. The mesh density has to be higher than that of physical nodes belonging to the liquid structure. Moreover, the set of the nodes of the liquid structure has to be a subset of the nodes of the CFD mesh. In other words, for each point $P_{in}$ there exists a point $P_m$ such that $P_{in} = P_m$; the same condition is required for the inlet and the outlet points $P_{inlet}$ and $P_{outlet}$, respectively.

The boundary conditions of the CFD problem are:
\begin{itemize}
\item wall condition: $v_i=0$ for each $P_m$ belonging to the boundary of $\Omega$ except for segments $I$ and $O$, where $v_i$ is the vector fluid velocity; 
\item inlet condition: normal velocity component $v_{inlet}$ at each point $P_m$ belonging to $I$; 
\item outlet condition: normal velocity component $v_{outlet}$ at each point $P_m$ belonging to $O$. 
\end{itemize}

Velocities $v_{inlet}$ and $v_{outlet}$ are set in such a way that they are linearly related to the liquid structure displacements, namely
\begin{equation} \label{ki}
   v_{inlet \; j}=k \, \delta_{inlet \,j}  \: \: (j\in\{1,...,n_{inlet}\})
\end{equation}
and 
\begin{equation} \label{ko}
   v_{outlet \; j}=k \, \delta_{outlet \,j} \: \: (j\in\{1,...,n_{outlet}\}),
\end{equation}
where $k$ is a constant to be chosen arbitrarily.

On the one hand, Eqs. \eqref{ki} and \eqref{ko} provide the velocity boundary conditions for just a subset of inlet and outlet $P_m$ points since the physical node density is lower than the density of the mesh. On the other hand, to solve the CFD problem, a complete set of inlet and outlet boundary velocities is required at all nodes of the mesh. In other words, values for $v_{inlet}$ and $v_{outlet}$ have to be provided for each point $P_m$ belonging to the boundary.

\begin{figure}[h!]
\centering
    \includegraphics[width=0.8\textwidth]{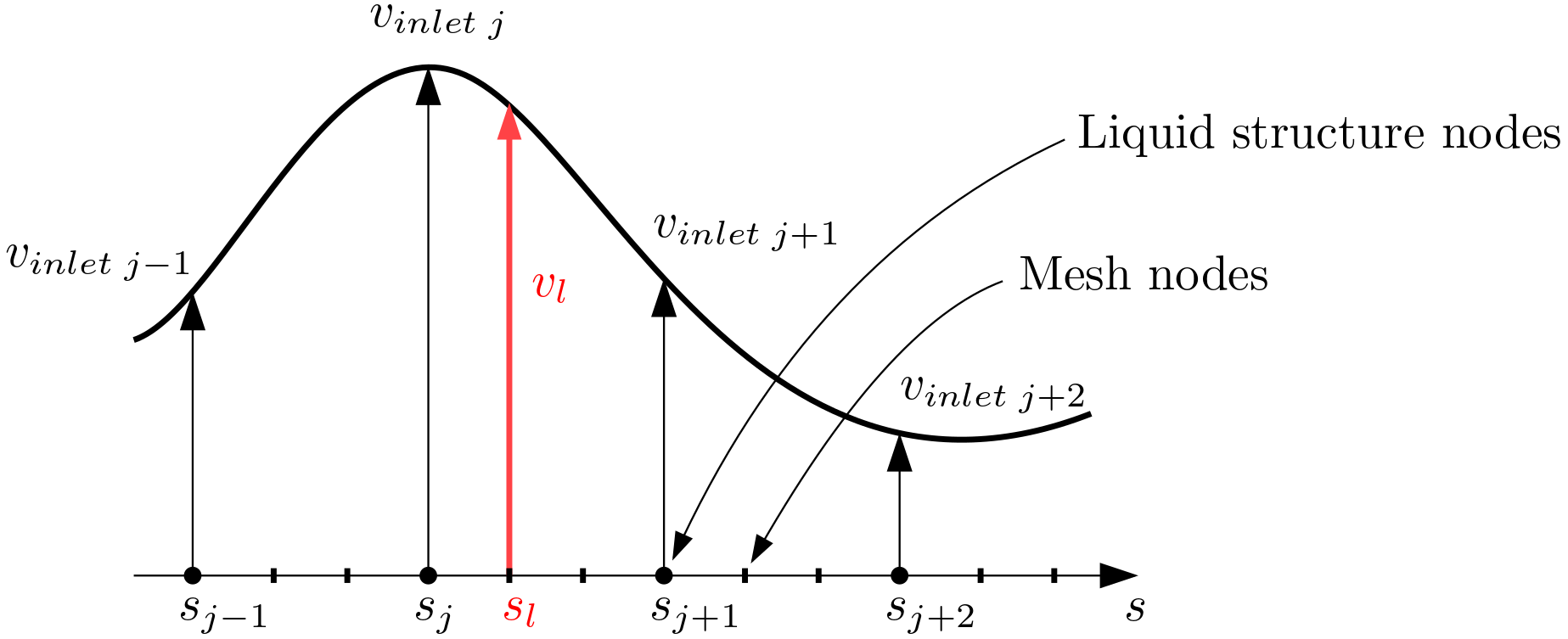}
    \caption{Interpolation of the inlet velocities in the inlet segment $I$.}
    \label{fig:interpolazione}
\end{figure}

One possibility to achieve the goal consists in interpolating the values of the boundary velocity in between two consecutive cell nodes as shown in Fig. \ref{fig:interpolazione}.

Referring to Fig. \ref{fig:interpolazione}, let us assume that $v_{inlet \; j-1}$, $v_{inlet \; j}$ and $v_{inlet \; j+1}$ are three consecutive boundary velocities calculated by means of Eqs. \eqref{ki} and \eqref{ko}. The coordinate $s$ is the 1-D local reference frame with the origin on the node where $v_{inlet \; j}$ is applied to. Therefore, the positions of these liquid structure nodes on the $<s>$ frame are respectively $s_{j-1}$, $0$ and $s_{j+1}$. The other boundary velocities for the mesh nodes can be interpolated by a cubic function, namely 
\begin{equation} \label{poly}
  v= \prod (s)\overset{\underset{\mathrm{def}}{}}{=}a_0+a_1 s+a_2 s^2 + a_3 s^3,
\end{equation}
where the constants $a_0, a_1, a_2$ and $a_3$ are computed by imposing the following constraints:
\vspace{-0.8cm}
\begin{multline}
\\ \prod (s_j)=v_{inlet \; j}, \\
\prod (s_{j+1})=v_{inlet \; j+1}, \\
\frac{\mathrm{d} }{\mathrm{d} s}\prod (s_j)=\frac{a_{inlet \; j}+a_{inlet \; j+1}}{2},  \\
\frac{\mathrm{d} }{\mathrm{d} s}\prod (s_{j+1})=\frac{a_{inlet \; j+1}+a_{inlet \; j+2}}{2},\\
\end{multline}
where $a_{inlet \; j}=(v_{inlet \; j}-v_{inlet \; j-1})/(s_j-s_{j-1})$.

It turns out that the interpolated velocity for the mesh border nodes are given by 

\begin{equation} \label{funzione_interpolante}
  v_l= \prod (s_l),
\end{equation}
where $s_l$ represents the mesh node position along the $<s>$ frame.
A similar interpolation procedure has to be applied to the outlet boundary velocities $v_{outlet \; j}$.
When $j=1$ or $j=n_{inlet}$, the conditions on $a_{inlet \; j}$ can not be applied. For these cases it is possible to impose the constraint 
\begin{equation}
\frac{\mathrm{d} }{\mathrm{d} s}\prod (s_j)=0.
\end{equation}
A cubic polynomial function is chosen since it allows one to impose constraints on the first-order derivative of the boundary velocities for the mesh nodes. The method could be further extended to a higher-order polynomial function, as for instance five-degree law, as well as to spline functions. In these cases, $C^2$ continuity at the interpolating points can be achieved.

If the CFD analysis is performed with an incompressible fluid, the continuity equation ($\bigtriangledown \cdot v = 0$) has to be guaranteed: the total inlet flow has to match the outlet one, namely
\begin{equation} \label{continuity}
  \int_{I} v_{inlet \; \perp}\: ds=\int_{O} v_{outlet \;\perp}\: ds,
\end{equation}
where the symbol $\perp$ refers to the velocity component perpendicular to the boundary.
If the displacements $\delta_{inlet}$ and $\delta_{outlet}$ are chosen in such a way to be perpendicular to the boundaries and discretization is considered, Eq. \eqref{continuity} becomes
\begin{equation} \label{costanti}
  \frac{\sum_{j=1}^{n_{inlet}}v_{inlet \; j}}{n_{inlet}}=  \frac{\sum_{j=1}^{n_{outlet}}v_{outlet \; j}}{n_{outlet}}.
\end{equation}

By using Eqs. \eqref{ki} and \eqref{ko} into \eqref{costanti}, it is possible to carry out the relationship between the inlet and the outlet displacements 
\begin{equation} \label{sommatoria}
  \frac{\sum_{j=1}^{n_{inlet}}\delta_{inlet \; j}}{n_{inlet}}=  \frac{\sum_{j=1}^{n_{outlet}}\delta_{outlet \; j}}{n_{outlet}}.
\end{equation}
Therefore, the inlet and the outlet displacements can not be chosen arbitrarily, at least if an incompressible fluid is employed.

Now that all the boundary conditions are defined, it is possible to carry out the CFD analysis using available CFD numerical tools. The results of the numerical analysis is the set of velocities $v_{in \; j} \; (j\in\{1,...,n_{in}\})$ for each point $P_{in \,j}$.
Given the velocities, the displacement assigned to each node is calculated by the relationship
\begin{equation} \label{delta_inverso}
  \delta_{in \,j}=1/k \; v_{in \; j}    \:\: (j\in\{1,...,n_{in}\}).
\end{equation}

\subsubsection{STEP 2: cell kinematic synthesis}
Now that the displacement of each cell node is assigned, the kinematic synthesis for each cell can be performed.

A liquid structure can be made up of different cells possessing different number of links and joints, and topology. In the following, to show the flexibility of the liquid structure philosophy, two different kinds of cells are introduced, namely  \textit{star cell} and \textit{diamond cell}.

\begin{figure}[h!]
\centering
    \includegraphics[width=0.6\textwidth]{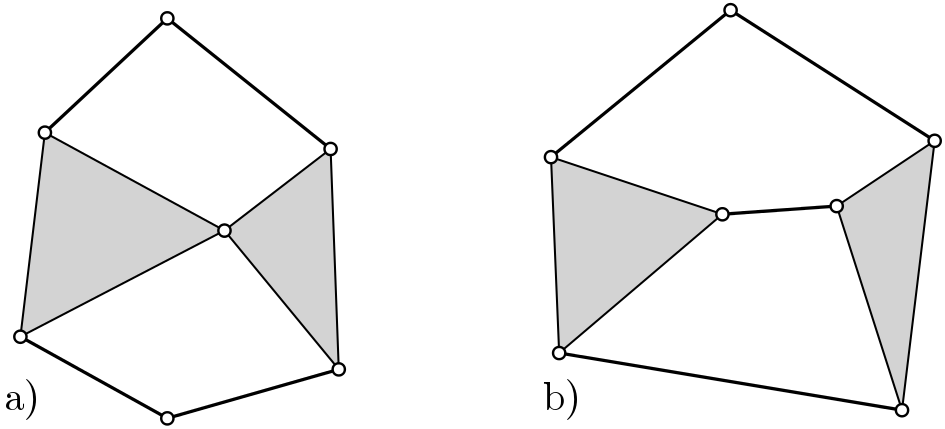}
    \caption{Possible layout of cell linkages: a) Watt's chain, b) Stephenson's chain.}
    \label{fig:linkages}
\end{figure}

As an alternative, other cell layouts that can be adopted, all based on the well known linkage topologies. For example, good candidates are either Watt linkages of kind I, II \cite{10.1115/1.1334346} or Stephenson I, II and III \cite{design_linkage}. Such linkages are based on the Watt and Stephenson chains (see Fig. \ref{fig:linkages}).

\subsection{Star cell}
\begin{figure}[h!]
\centering
    \includegraphics[width=0.6\textwidth]{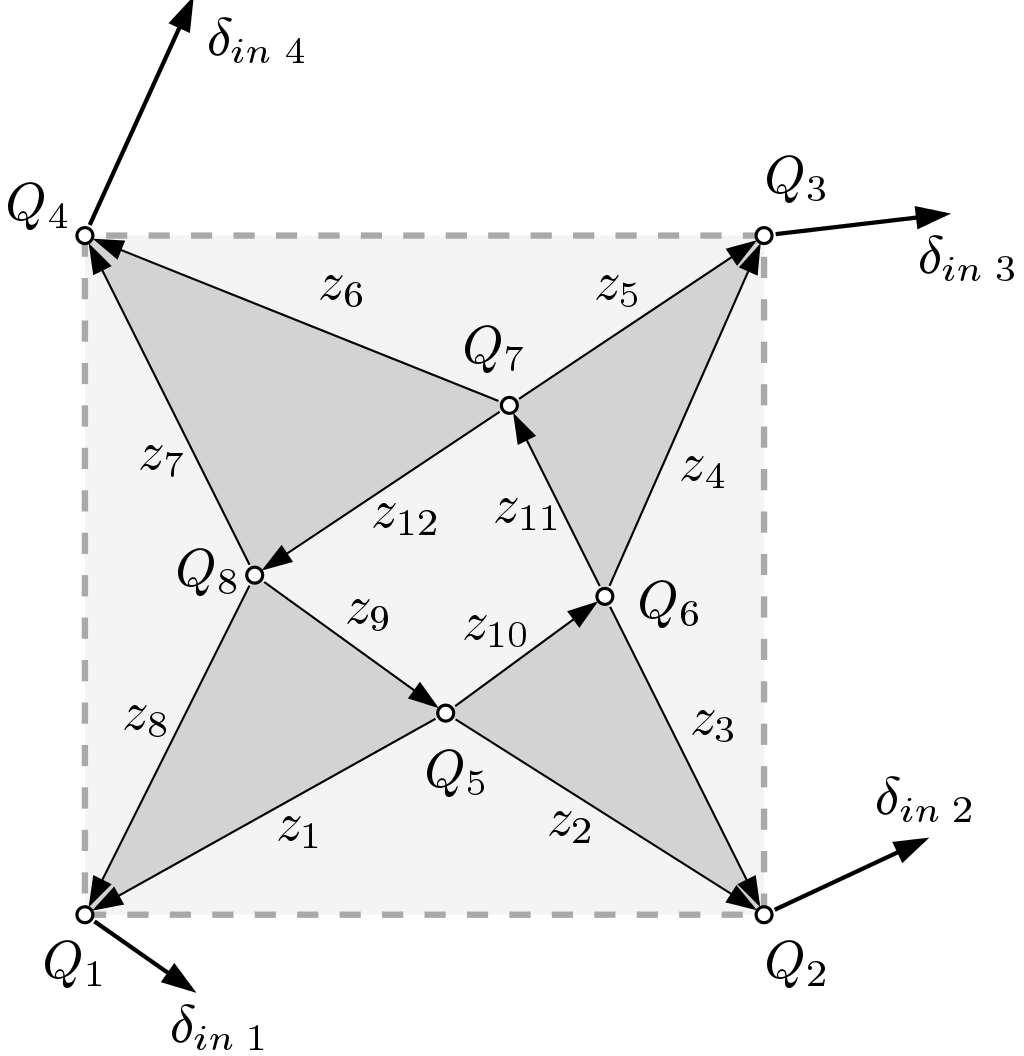}
    \caption{Sketch of a star cell and relevant nomenclature.}
    \label{fig:cell_linkage}
\end{figure}

Figure~\ref{fig:cell_linkage} represents the sketch of a 4-node star cell. Points \(Q_{1},\, Q_{2},\, Q_{3}\) and \(Q_{4}\) are the nodes of the cell. To simplify the notation, the subscript that relates the points to the cell $j$-th has not been displayed. Moreover, the point coordinates are expressed with respect to a local reference frame with the origin in $Q_1$. Basically, the cell is made up of a 4-bar linkage connected to 4 rigid triangles. It has 4 DoF, since 
\begin{equation} 
   n_{DoF_j}=3\, nl_{j}-2\, nc_j=3\times12-2\times16=4. 
\end{equation}

Without losing any generality, the star cell can be of dimensions such that it can be inscribed in a square of unitary side. A complex variable notation is adopted to describe the kinematics: real and imaginary values are associated with $x$ and $y$ coordinates, respectively. Therefore, the coordinates of the points with respect to the local reference frame are $Q_{1}=0,\ Q_2=1,\ Q_3=1+i$ and $Q_4=i$.

Vector $z_k$ is aligned along the rigid link. By introducing the symbols $h:=1$ and $u:=i$, the geometrical kinematic constrains for the star cell in its initial configuration are:
\begin{multline} \label{s1}
\\ -z_1+z_2=h \\
-z_3+z_4=u \\
-z_6+z_5=h \\
-z_8+z_7=u \\
\end{multline}
and, after introducing the displacements:
\begin{multline} \label{s2}
\\ -\delta_{in\;1}+h+\delta_{in\;2}=-z_1\;e^{\Delta\;\beta_1}+z_2\;e^{\Delta\;\beta_2} \\
-\delta_{in\;2}+u+\delta_{in\;3}=-z_3\;e^{\Delta\;\beta_2}+z_4\;e^{\Delta\;\beta_3} \\
-\delta_{in\;3}-h+\delta_{in\;4}=-z_5\;e^{\Delta\;\beta_3}+z_6\;e^{\Delta\;\beta_4} \\
-\delta_{in\;4}-u+\delta_{in\;1}=-z_7\;e^{\Delta\;\beta_4}+z_7\;e^{\Delta\;\beta_1}. \\
\end{multline}

The notation $\Delta\;\beta_1$ represents the angular rotation of the rigid triangle $Q_1,\ Q_5,\ Q_8$ that occurs when the cell nodes moves from locations $Q_1,\ Q_2,\ Q_3,\ Q_4$ to locations $Q_1+\delta_{in \,1},Q_2+\delta_{in \, 2},Q_3+\delta_{in \,3},Q_4+\delta_{in \,4}$.
Note that $\delta_{in \,1}$, $\delta_{in \, 2}$, $\delta_{in \,3}$ and $\delta_{in \,4}$ are the node displacements obtained at the end of \textit{STEP 1}.

The quantity $z_1\,e^{\Delta\;\beta_1}$ represents the vector $z_1$ after the rigid rotation. In other words, $e^{\Delta\;\beta_1}$ plays the role of an angular operator. Same considerations apply for the other angular rotation parameters $\Delta\;\beta_k$.   

\begin{figure}[h!]
\centering
    \includegraphics[width=0.6\textwidth]{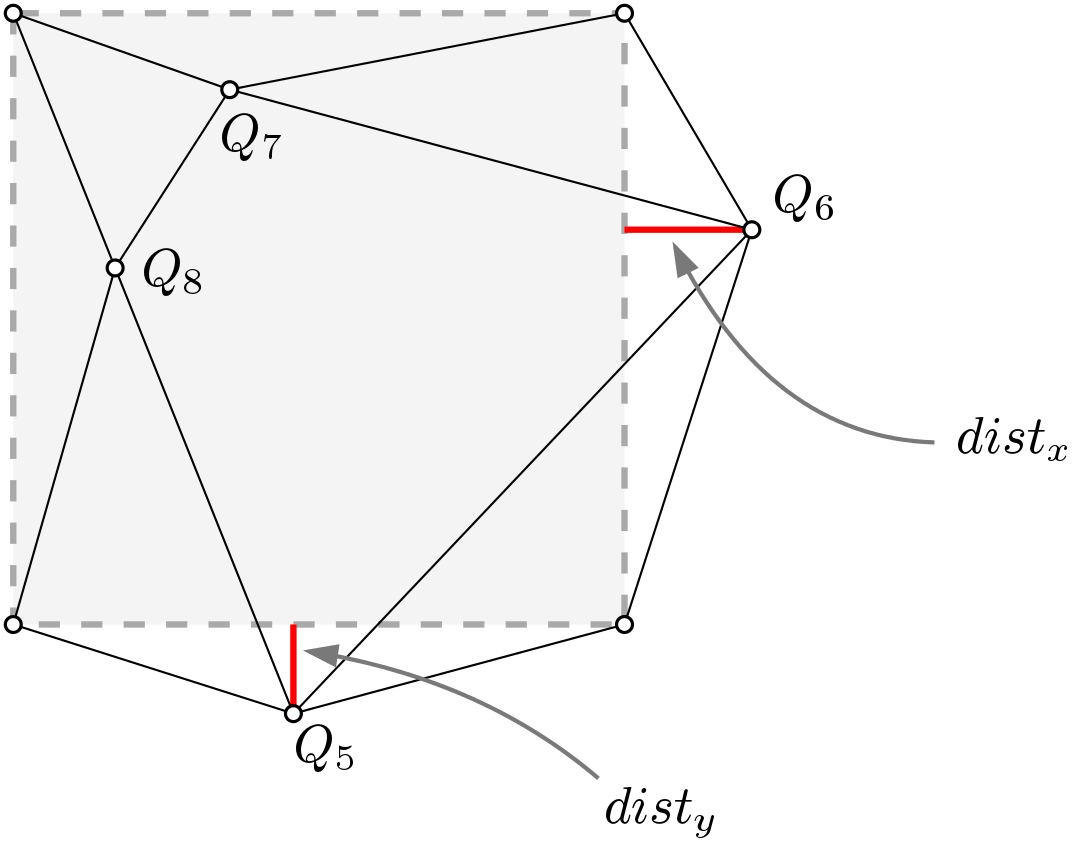}
    \caption{Graphical representation of $\text{dist}_x(Q_l)$ and $\text{dist}_y(Q_l)$ for nodes $Q_6$ and $Q_5$, respectively.}
    \label{fig:stella_distanze}
\end{figure}

In matrix notation, the combination of Eqs. \eqref{s1} and \eqref{s2} leads to the system
\begin{equation} \label{sistema}
\boldsymbol{A}_s\boldsymbol{z}_s=\boldsymbol{u}_s,
\end{equation}
where $\boldsymbol{A}_s$, $\boldsymbol{z}_s$ and $\boldsymbol{u}_s$ are defined in \ref{matrices}.
The system has 8 complex unknowns ($z_1,z_2,z_3,z_4,z_5,z_6,z_7,z_8$) and 4 scalar unknowns ($\Delta\;\beta_1,\Delta\;\beta_2,\Delta\;\beta_3,\Delta\;\beta_4$). Therefore, it is underdetermined.
A linear solution of the system is provided, but it is a function of the parameters $\Delta\;\beta_1,\Delta\;\beta_2,\Delta\;\beta_3,\Delta\;\beta_4$, namely
\begin{equation}
    \boldsymbol{z}_s=\boldsymbol{A}_s^{-1}\boldsymbol{u}_s=f(\Delta\;\beta_1,\Delta\;\beta_2,\Delta\;\beta_3,\Delta\;\beta_4).
\end{equation}
The four parameters can be chosen so as to minimize a cost function related to the kinematic configuration of the cell. In particular, the following cost function is introduced:
\begin{equation}
   f_C=f_C(\Delta\;\beta_1,\Delta\;\beta_2,\Delta\;\beta_3,\Delta\;\beta_4)=w_1+w_2,
   \label{coststar}
\end{equation}
where $w_1=1/\left \| \det\boldsymbol{A}_s \right \|$ and $w_2=\text{dist}(Q_5)+\text{dist}(Q_6)+\text{dist}(Q_7)+\text{dist}(Q_8)$. In the latter, $\text{dist}(Q_l)=\text{dist}_x(Q_l)+\text{dist}_y(Q_l)$, where

\begin{equation}
\text{dist}_x(Q_l)=\left\{\begin{matrix}
0 \ \  \textrm{if} \ \  0\leq \text{real}(Q_l)\leq 1
\\ 
- \text{real}(Q_l)  \ \ \textrm{if} \ \  \text{real}(Q_l) <0
\\
 \text{real}(Q_l)-1  \ \ \textrm{if} \ \  \text{real}(Q_l) >1
\end{matrix}\right.
\end{equation}
and

\begin{equation}
\text{dist}_y(Q_l)=\left\{\begin{matrix}
0 \ \  \textrm{if} \ \  0\leq \text{imag}(Q_l)\leq 1
\\ 
- \text{imag}(Q_l)  \ \ \textrm{if} \ \  \text{imag}(Q_l) <0
\\
 \text{imag}(Q_l)-1  \ \ \textrm{if} \ \  \text{imag}(Q_l) >1
\end{matrix}\right.
\end{equation}
On the one hand, the presence of weight  $w_1$ in the cost function \eqref{coststar} prevents the matrix $\boldsymbol{A}_s$ from becoming singular at the solution; on the other hand, $w_2$ tends to find solutions for which the points $Q_1,Q_2,Q_3,Q_4$ do not exit the unitary square of the star cell. This is desirable in order to avoid that mechanisms of two adjacent cells interfere with each other (outside the unitary side square) (see Fig. \ref{fig:stella_distanze}).

\subsection{Diamond cell}
\begin{figure}[h!]
\centering
    \includegraphics[width=0.6\textwidth]{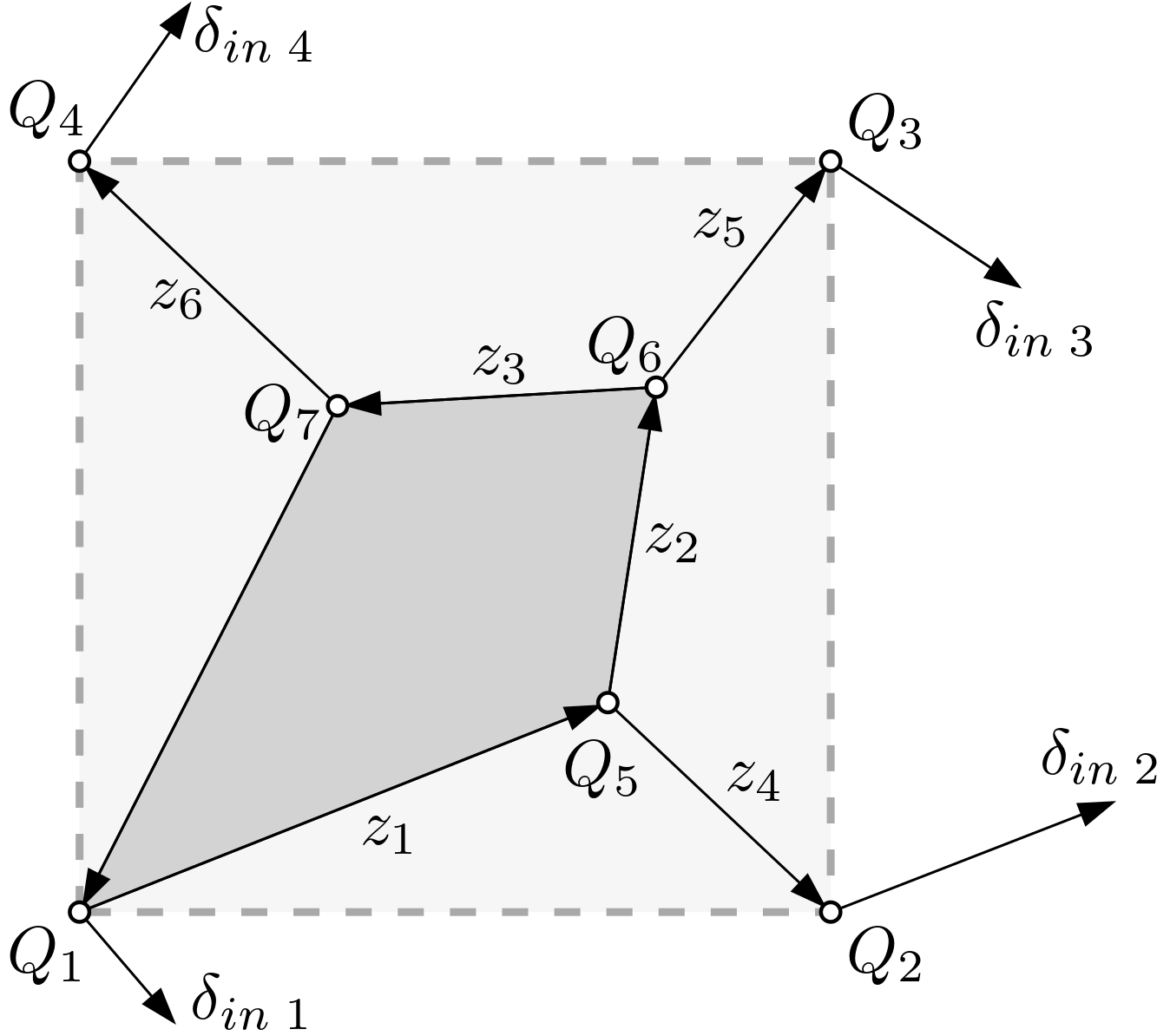}
    \caption{Sketch of a diamond cell and relevant nomenclature.}
    \label{fig:diamante}
\end{figure}

In Fig.~\ref{fig:diamante} the sketch of a diamond cell is shown. \(Q_{1}, Q_{2}, Q_{3}\) and \(Q_{4}\) are the nodes of the cell. As for the previous cell, the subscript that relates the points to the $j$-th cell has not been displayed and the point coordinates are expressed in a local reference frame with the origin in $Q_1$. The cell is made up of a rigid irregular quadrilateral connected to the three nodes by means of  rockers. It has 6 DoF, since (the revolute joints on the nodes are not considered since they are used just to connect the cell to the adjacent ones)
\begin{equation} 
   n_{DoF_j}=3\, nl_{j}-2\, nc_{j}=3\times4-2\times3=6 .
\end{equation}

Also in this case, the  cell  dimensions are such that it can be inscribed within a square of unitary side and in the initial configuration, the coordinates of the points  are $Q_{1}=0,Q_2=1,Q_3=1+i,Q_4=i$.

The geometrical kinematic constraints for the diamond cell in the initial configuration are:
\begin{multline} \label{ss1}
\\ z_1+z_4=h \\
-z_4+z_2+z_5=u \\
-z_6-z_3+z_5=h \\
\end{multline}
that, after having made explicit the displacement, can be written as:
\begin{multline} \label{ss2}
\\ -\delta_{in\;1}+h+\delta_{in\;2}=z_1\;e^{\Delta\;\beta_1}+z_4\;e^{\Delta\;\beta_4} \\
-\delta_{in\;2}+u+\delta_{in\;3}=-z_4\;e^{\Delta\;\beta_4}+z_2\;e^{\Delta\;\beta_1}+z_5\;e^{\Delta\;\beta_5} \\
-\delta_{in\;3}-h+\delta_{in\;4}=z_6\;e^{\Delta\;\beta_6}+z_3\;e^{\Delta\;\beta_1}-z_5\;e^{\Delta\;\beta_5}. \\
\end{multline}

The phase $\Delta\;\beta_1$ represents now the angular rotation of the rigid quadrilateral $Q_1,Q_5,Q_6,Q_7$ that takes place when the  nodes move from locations $Q_1,Q_2,Q_3,Q_4$ to $Q_1+\delta_{in \,1},Q_2+\delta_{in \,2},Q_3+\delta_{in \,3},Q_4+\delta_{in \,4}$, respectively.  
In matrix notation, the combination of Eqs. \eqref{ss1} and \eqref{ss2} leads to the system (see \ref{matrices})
\begin{equation} \label{sistema2}
\boldsymbol{A}_d\boldsymbol{z}_d=\boldsymbol{u}_d.
\end{equation}

For the diamond cell
the system has 6 complex unknowns ($z_1,z_2,z_3,z_4,z_5,z_6$) and 4 scalar unknowns ($\Delta\;\beta_1,\Delta\;\beta_4,\Delta\;\beta_5,\Delta\;\beta_6$), and is, therefore, underdetermined.
A linear solution of the system is provided, but is a function of the four parameters listed in the latter bracket, namely
\begin{equation}
    \boldsymbol{z}_d=\boldsymbol{A}_d^{-1}\boldsymbol{u}_d=f(\Delta\;\beta_1,\Delta\;\beta_4,\Delta\;\beta_5,\Delta\;\beta_6).
\end{equation}

Similarly to the star cell, the four parameters are chosen to minimize the cost function 
$
   f_C=w_1+w_2,
$
where $w_1=1/\left \| \det\boldsymbol{A}_d \right \|$ and $w_2=\text{dist}(Q_5)+\text{dist}(Q_6)+\text{dist}(Q_7)$.

It should be emphasised that both star and diamond cells represent just two of the possible topologies that could be employed to assemble a liquid structure. We believe that depending on the type of structure at hand,
there could exist more effective cells than those we proposed. To this regard, as it will be pointed out in Section \ref{future}, several aspects should be explored and taken into account in order to select the proper cell, such as the energy required to allow soft modes, the occurring of either singular configurations  or link crossing.  The selection and study of the properties of such optimized cells represent an interesting future research field since the linkage combinations are numerous and not always the kinematic properties are easy to predict.  

\subsection{Liquid structures are made of bistable mechanisms}
A mechanism is bistable when it has two different stable equilibrium configurations. A cell of a liquid structure is conceived to be a bistable mechanism. The first equilibrium configuration, say $C_1$, corresponds to the reference one, with cell nodes located at points $Q_j$. The second equilibrium configuration, say $C_2$, is reached when nodes are at positions $Q_j+\delta_j$. The common ground is that in these two configurations all the link lengths are the same, however, during the transition from one configuration to the other, the structure must exhibit a certain degree of compliance in order to compensate for the momentary closeness and relocation of the nodes.

\begin{figure}[h!]
\centering
    \includegraphics[width=0.95\textwidth]{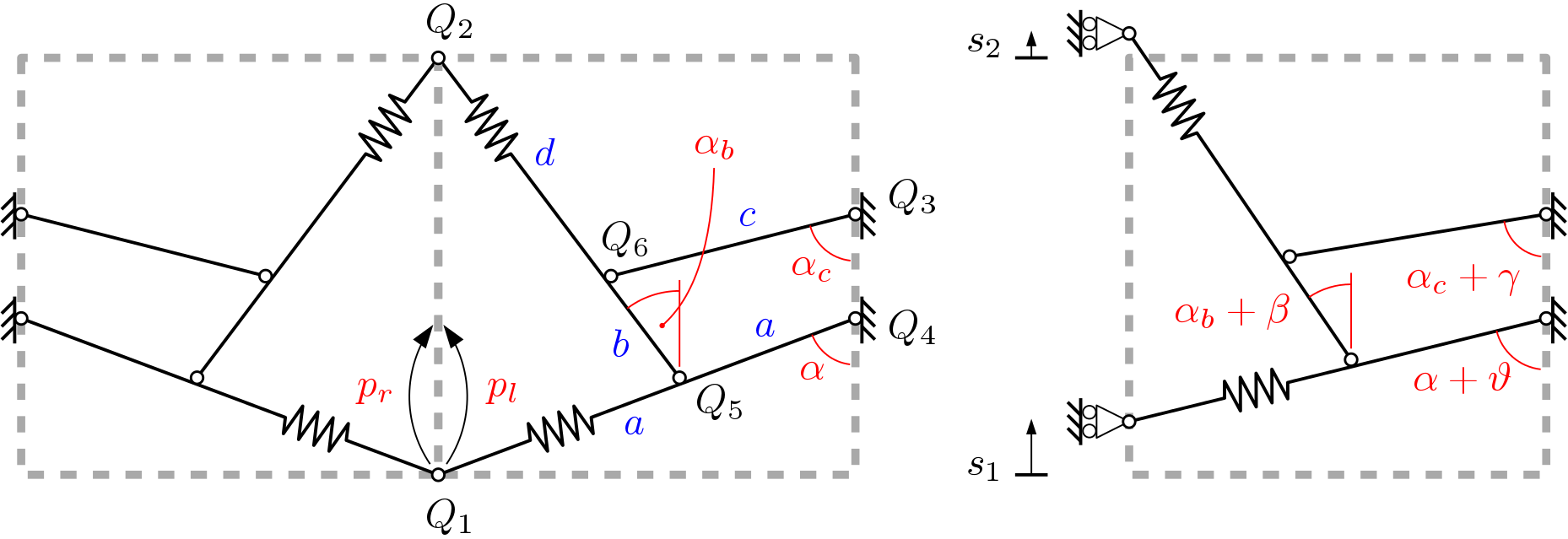}
    \caption{Analysis of bistability of two symmetric cells composed of three pseudo-rigid links. Each cell occupies a square whose side is unitary. Left: reference configuration ($C_1$). Right: generic deformed configuration; angles $\beta$ and $\gamma$ and displacements $s_1$ and $s_2$ are functions of the degree of freedom $\theta$. Links $Q_1 Q_5$ and $Q_2 Q_6$ are axially linear elastic.}
    \label{fig:bistable}
\end{figure}

Figure \ref{fig:bistable} exemplifies the concept. For the sake of simplicity, instead of two star/diamond cells, a simpler topology is introduced,  composed of three pseudo-rigid links and possessing one DoF, i.e. the angle $\theta$.
Let us consider the point $Q_1$. If the two cells  were decoupled, to get $C_2$ from $C_1$ the point $Q_1$, considered as an element of the cell on the left, would move along the path $p_l$. On the contrary, the same point, seen from the cell on the right, would move along the path $p_r$. As a consequence, when the two cells are connected, symmetry reasons suggest that the actual path is a straight segment. It is clear that each cell must have some compliant elements in order to allow link shortenings.

An equilibrium and stability investigation can be performed assuming the least number of elastic links to illustrate how the mechanism can switch from $C_1$ to $C_2$. We assume that links $Q_1 Q_5$ and $Q_2 Q_6$ are axially linear elastic with stiffness $K$ while rigid under both bending and shear. The cell occupies the square whose side is unitary and is designed to achieve $\delta_1=2 \delta_2=0.4$. Upon application of an upward vertical load $F$ to $Q_1$ (other loading conditions respecting the symmetry can also be imposed to the mechanism), the cell displaces and the elastic links shorten. With reference to the detailed information reported in Appendix B, the total potential energy $W(\theta)$ of the cell takes the form
\begin{equation}
    W (\theta)=\frac{1}{2} K \eta^2_1(\theta)+\frac{1}{2} K \eta^2_2(\theta)-F s_1(\theta),
\end{equation}
where $\eta_1(\theta)$ and $\eta_2(\theta)$ represent the axial deformations of the two springs and $s_1(\theta)$ the displacement of point $Q_1$.  
Equilibrium configurations correspond to solution  $\bar\theta$ such that $W'(\theta)_{|\bar \theta}=0$. Three solutions are detected for $F=0$, namely the two with the links undeformed, i.e. $\bar\theta=0$, $\bar\theta=0.3948$, and the intermediate one, $\bar\theta=0.1995$, in which the elastic links are compressed, but equilibrium of the cell is still guaranteed. While the first two are stable configurations, i.e. $W''(\theta)>0$, the third one turns out to be unstable. 

The foregoing analysis is typical of snap-through structures, however if the kinematics of the cell required more than one DoF the energy landscape would become much more complex, but still allowing the two zero-strain energy stable modes obtained from the synthesis.

\subsection{Degrees of freedom of a cell}

Maxwell \cite{Maxwell1865} was one of the first that studied the stability of lattice  structures. In particular, he introduced the concept of \textit{minimum connectivity}. If we indicate with $h_n$ the average number of nodes each node is connected to, there exists a value $h_{nl}$ below which the
structure displays soft modes or, in other words, deformations at zero cost energy. For a large 2D network composed of $N$ nodes, the critical connectivity is $h_{nl}=4$. This result was obtained by balancing the number of translational DoF of all nodes ($2N$) with the number of constraints or bonds.

Following a similar approach (still based on balancing the overall number of DoF with internal and boundary constraints) a new structure stability criterion is presented. In our case, the stability is guaranteed in terms of cell \#DoF.  

In the previous two Subsections, we introduced two different cells: the star cell and the diamond one, with 4 and 6 DoF, respectively. For a liquid structure it is impossible to have all cells with less than 4 DoF (3 DoF or less) because they would not be able to deform. In fact, a cell with just 3 DoF has zero internal mobility, or equivalently, it is a simple rigid body, and a liquid structure made up of just rigid bodies can not display soft modes.

Less evident is the largest \#DoF the cells can have. To answer this question we propose the following theorem:
\begin{itemize}
\item[ ] Given a rectangular liquid structure made up of cells with identical topology, there exists always a sufficiently large number of cells for which the liquid structure mobility is null (soft modes are not considered) when the wall nodes are fixed to the reference frame. Such claim is true if the \#DoF of each single cell is lower than 6.
\end{itemize}

\begin{proof}
Let be $r$ and $l$ the number of cells along the horizontal and the vertical axis, respectively (see Fig. \ref{fig:square_ls}). Let be $g$ the \#DoF of a single cell. Applying the Grubler equation to the entire liquid structure, and considering that the wall nodes are fixed to the reference frame by means to revolute joints, the total \#DoF of the liquid structure is
\begin{equation}
    LS_{DoF}=grl-2(4+2(l-1)+2(r-1)+3(l-1)(r-1)),
\end{equation}
where the product $grl$ represents the total \#DoF of all the cells without constraints (not connected to the others); $4$ are the corners, each of one has a revolute joint; $(l-1)$ and $(r-1)$ are the constraints on each side (excluding the corners) and $(l-1)(r-1)$ are the internal nodes, each one connects the 3 adjacent cells by means of 3 revolute joints.   

\begin{figure}[h!]
\centering
    \includegraphics[width=0.8\textwidth]{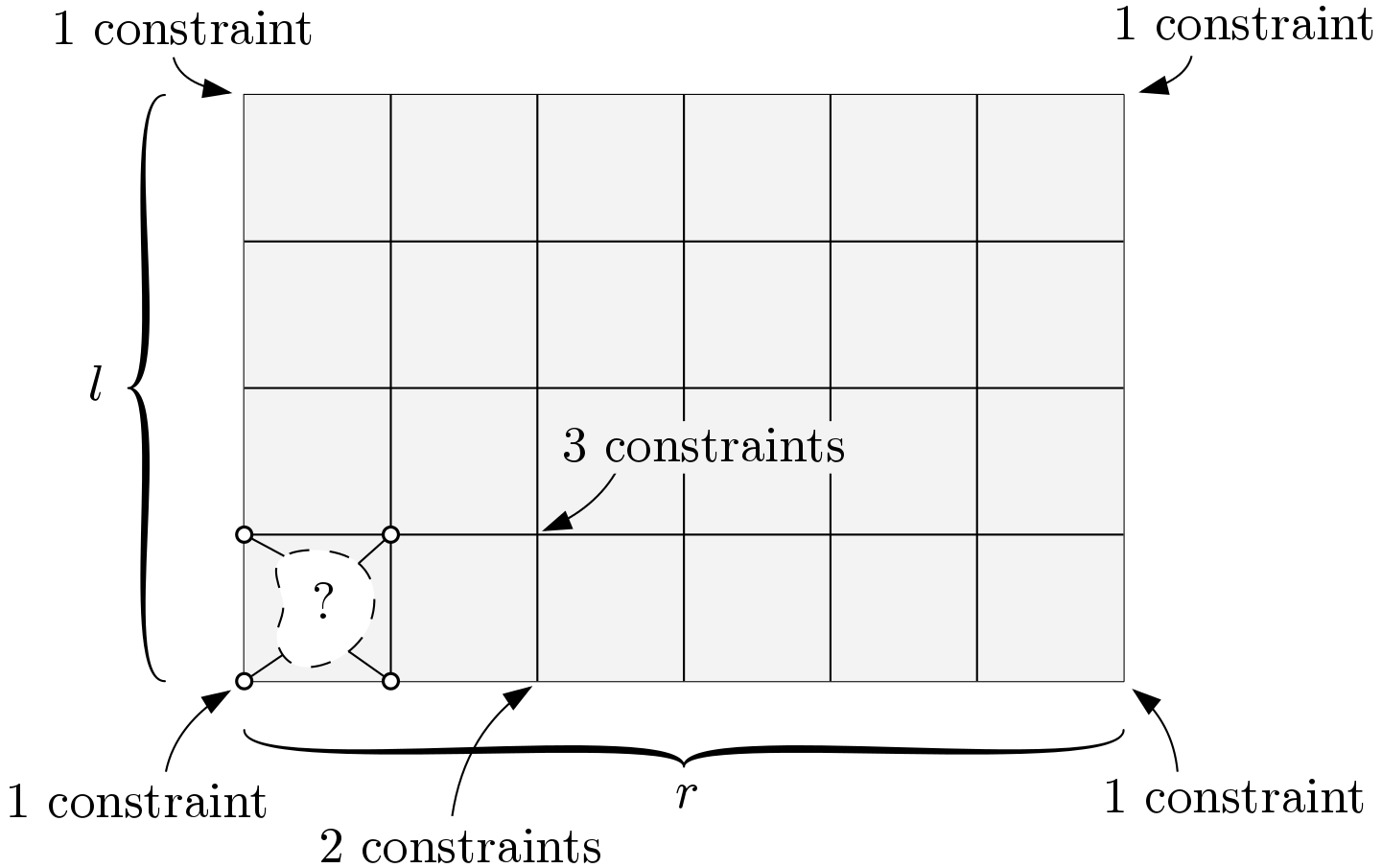}
    \caption{Schema related to a matrix of equal cells inside a rectangular area.}
    \label{fig:square_ls}
\end{figure}

The liquid structure has mobility null if $LS_{DoF}\leqslant0$. This condition leads to
\begin{equation}
    g\leqslant 6+\frac{6}{rl}-\frac{2}{r}-\frac{2}{l}.
\end{equation}
By taking the $\lim_{r,l\to\infty}$ of the right-hand side, the inequality becomes
\begin{equation}
    g\leqslant 6.
\end{equation}
\end{proof}

\section{An example of liquid structure} \label{implementation}

\begin{figure}[h!]
\centering
    \includegraphics[width=0.7\textwidth]{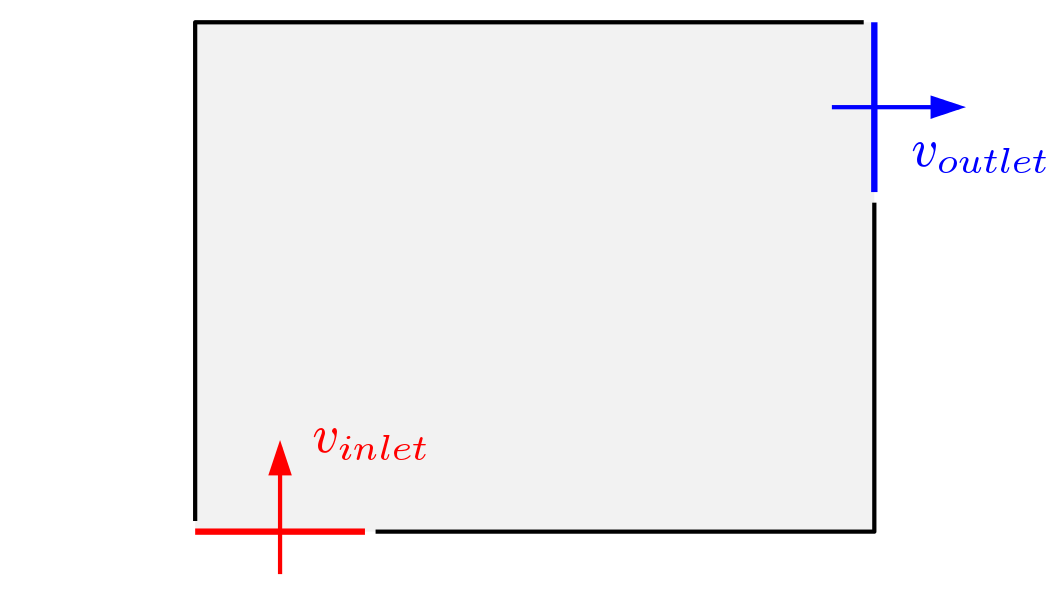}
    \caption{Rectangular domain ($15 \times 10$ in dimensionless units) of the example of liquid structure analysed via CFD where inlet and outlet boundaries, whose dimensionless lengths are 6 and 7, respectively, are displayed. }
    \label{fig:test}
\end{figure}

In order to show how to apply the described procedure, a prototype case is analysed and discussed.
The area of the simulated liquid structure is a rectangle whose dimensions, in non-dimensional units, are $15\times10$ (see Fig. \ref{fig:test} where the boundaries of the internal cells are not sketched). The goal is to create a soft mode such that small displacements imposed on the nodes of the inlet at the bottom left (sketched in red) produce a deformation of the outlet (sketched in blue). 

 The liquid structure can be ``filled'' with different cells; some of them could be star cells. We decided to locate the nodes of the cells $P_{in}$ at the corners of the unitary edge squares that tasselate the rectangular domain. Therefore, according to the protocol defined in STEP 1, the initial coordinates $P_j$ are given by the vertices of the $15\times10$ rectangular grid. The fluid boundary velocity $v$ is set to 0 at the wall nodes $P_{wall}$.
For the sake of simplicity, instead of using the cubic interpolation of Eq. \eqref{poly} as inlet boundary condition, a velocity $v_{inlet}=[0,1]$ is assigned to all the nodes of the inlet segment $P_{inlet}$.
The velocities of the outlet border have not been fixed \textit{a priori} at the points belonging to $P_{outlet}$. Instead, as it is customary using the features of the CFD software, they are set so as to satisfy the continuity equation ($\bigtriangledown \cdot v = 0$).

The software used to perform the CFD analysis is \textit{EasyCFD}, implemented at the University of Coimbra, Portugal \cite{Lopes201627}. Parameters set are density ($=1000$ kg/m$^3$), dynamic viscosity ($=10^{-3}$ Ns/m$^2$), pressure coefficient ($=4182$ J/kg) and Prandtl number ($=7.01$). The analysis is isothermal, the flow type is laminar and the regime is set to steady state. 

Instead of inlet velocity boundary conditions, it is possible to use the inlet pressure boundary conditions. Generally, pressure boundary conditions are used when the inlet pressure is known, but the flow rate and/or velocity is not known. This situation may arise in many practical fluid dynamics situations. In any case, the goal of our CFD analysis is not the simulation of a real case. It is just a numerical abstraction useful to generate a continuous field of displacements.

\begin{figure}[h!]
\centering
    \includegraphics[width=0.6\textwidth]{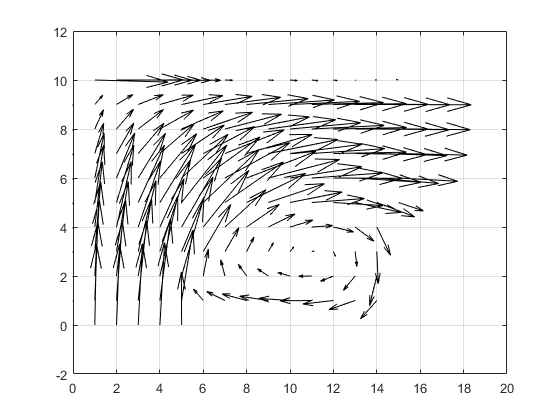}
    \caption{Velocity field computed by the CFD analysis. The inlet velocity is set as the vector $[0,1]$. }
    \label{fig:frecce}
\end{figure}

The velocity field computed by the CFD analysis for the selected geometry is shown in Fig. \ref{fig:frecce}. The convergence of the solution has been verified by mesh refining.

\begin{figure}[h!]
\centering
    \includegraphics[width=0.95\textwidth]{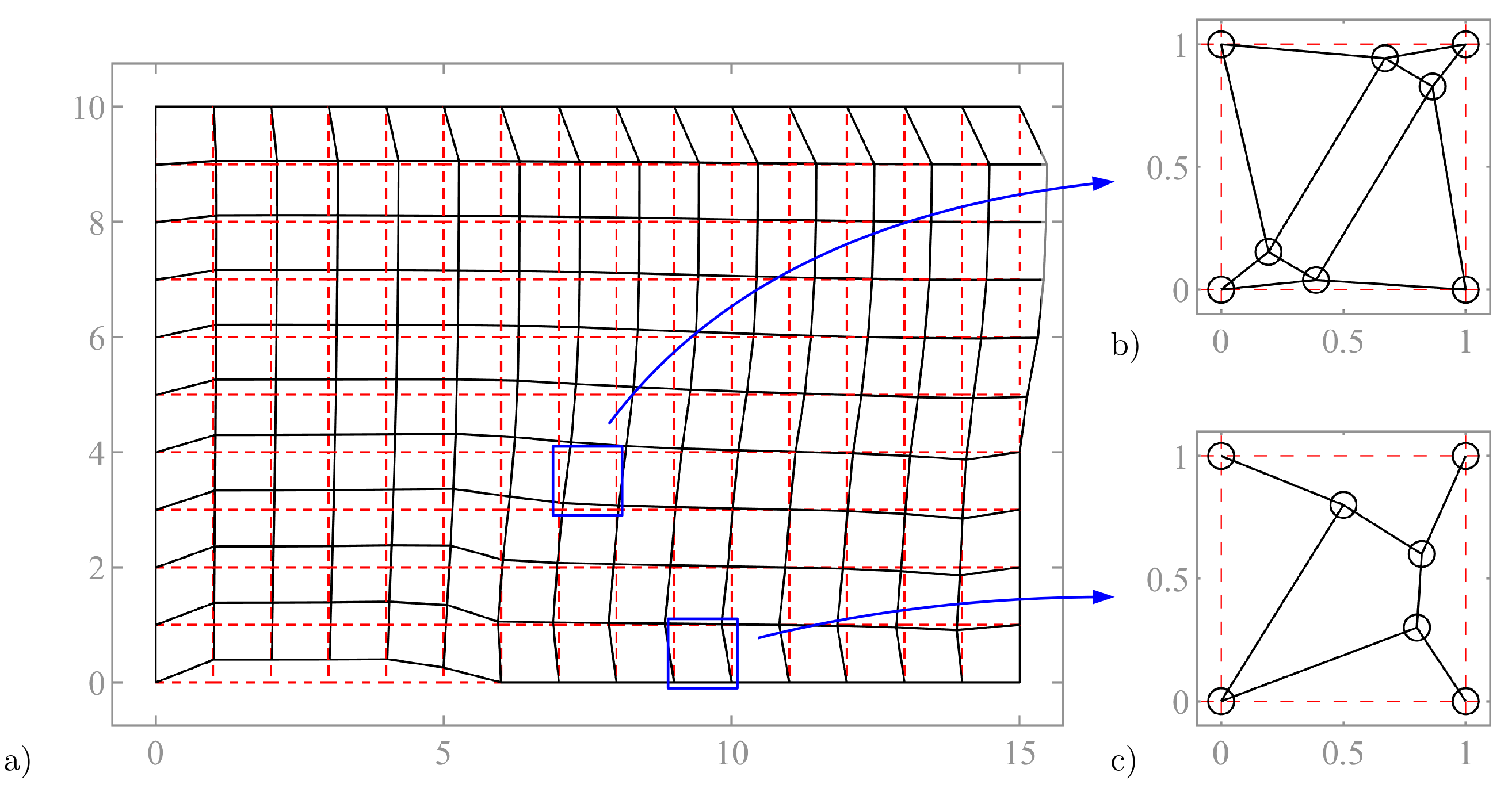}
    \caption{Liquid structure. In a) the unitary square cells making up the structure are shown; black solid lines refer to the cells at their initial positions; red dashed lines refer to the displaced configuration. A detailed view is shown for two cells located in the domain: b) a star cell and c) a diamond cell.}
    \label{fig:liquid_structure}
\end{figure}

Now that all the node velocities are known, the node displacements $\delta_{in}$ are calculated by means of Eq. \eqref{delta_inverso}, where $k$ was set to 0.4. Figure \ref{fig:liquid_structure} shows the cells (just the nodes) in the initial configuration (red dashed line) and the soft mode (black solid line).

\begin{figure}[h!]
\centering
    \includegraphics[width=0.6\textwidth]{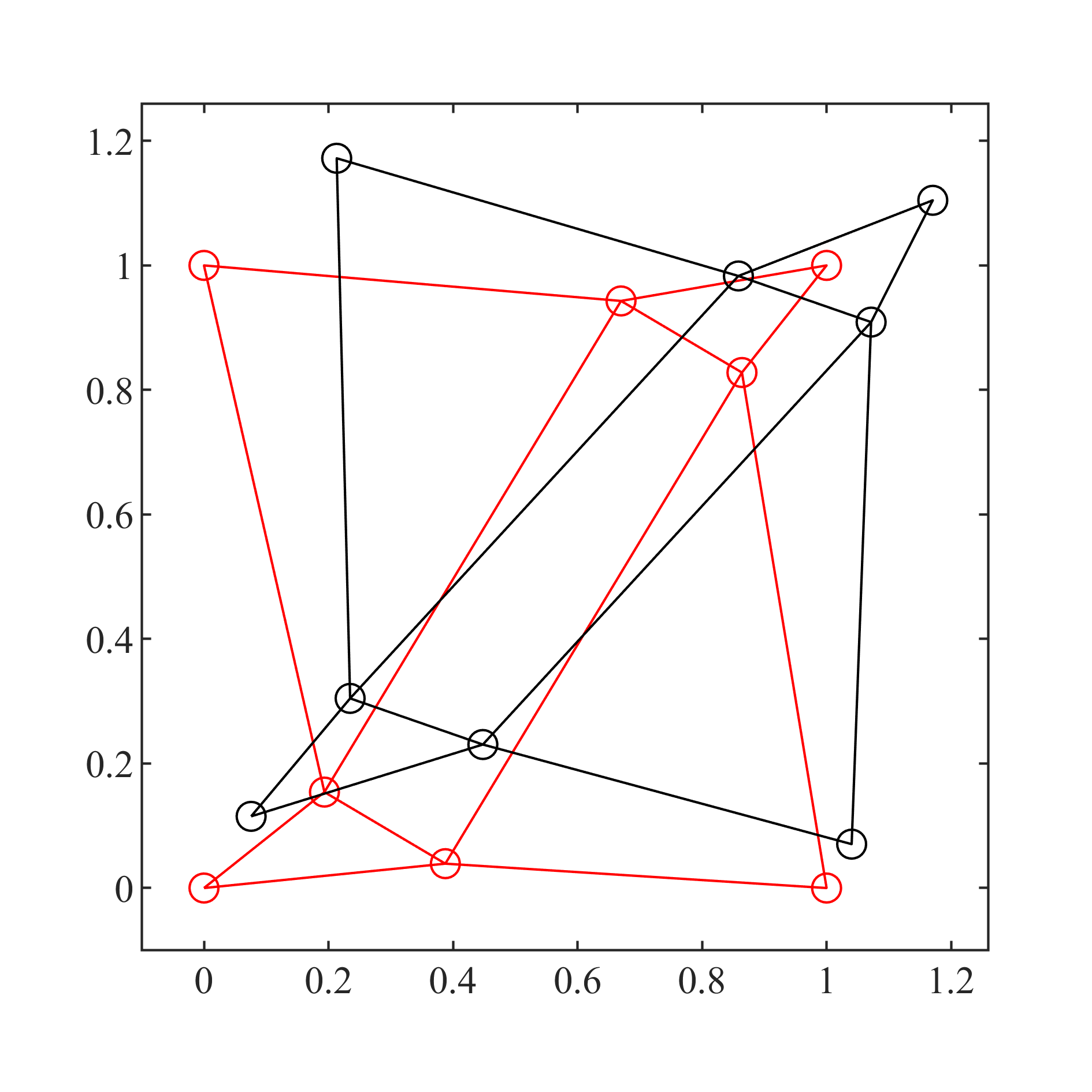}
    \caption{Sketch of the reference (red) and deformed (black) configurations of the star cell displayed in Fig. \ref{fig:liquid_structure}b).}
    \label{fig:starcell}
\end{figure}

Given the values of the internal node displacements, following STEP 2, the synthesis of each single cell can be performed. In Fig. \ref{fig:liquid_structure}a), the initial (dashed red line) and deformed (solid black line) configurations of the liquid structure are presented. As an example, the result of the synthesis of a star cell on position 8 (from the left) and 4 (from the bottom) is sketched in Fig. \ref{fig:liquid_structure}b). In Fig. \ref{fig:liquid_structure}c), the result of the synthesis of a diamond cell on position 10 (from the left) and 1 (from the bottom) is sketched as well. The deformed configuration of the former cell is also sketched in black in Fig. \ref{fig:starcell}  where it is evident that all four nodes initially in the corners of the square are subjected to displacements.

\section{A liquid structure implementation of a disk brake system }

\begin{figure}[h!]
\centering
    \includegraphics[width=0.6\textwidth]{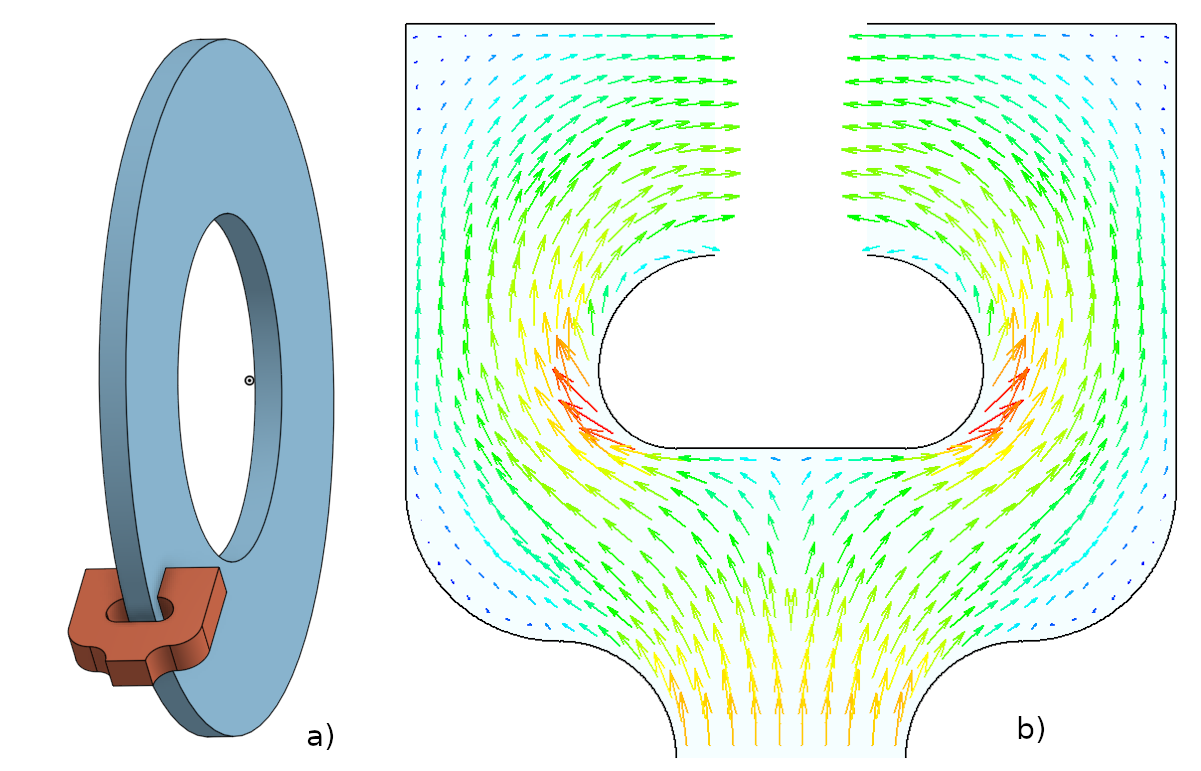}
    \caption{A disk brake system conceived as a liquid structure. (a) Sketch of the brake system. (b) Velocity profile computed by the CFD software.}
    \label{fig:rendering_velocita}
\end{figure}

\begin{figure}[h!]
\centering
    \includegraphics[width=0.6\textwidth]{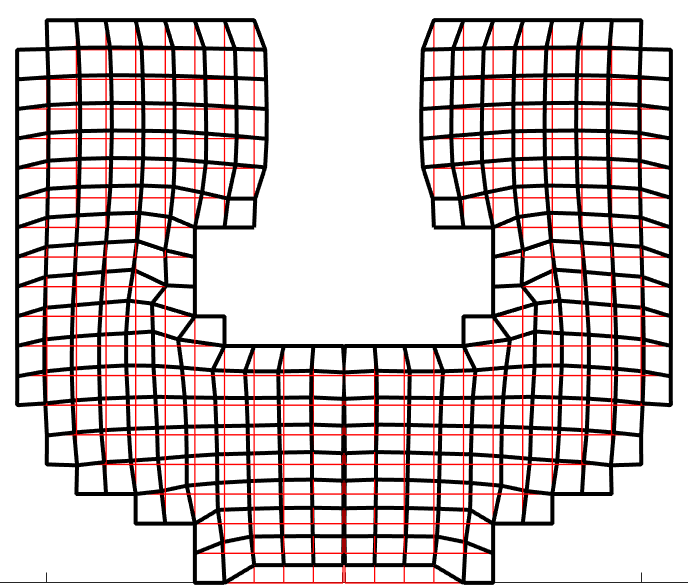}
    \caption{A disk brake system conceived as a liquid structure. Sketch of the arrangement of mechanism cells to ensure mechanical transmission and the designed braking performance. Red solid lines represent initial configurations of the cells; black solid lines represent the final configurations of the cells.}
    \label{fig:celle}
\end{figure}

To show the industrial exploitation potential of liquid structures, we propose the use of the  new paradigm in the design of a disk brake system. Usually the pads are squeezed against the disc through hydraulic actuation. A liquid structure can produce an equivalent squeezing action without the need of pistons and cylinders. 
The calipers are realized by a liquid structure mechanical fork (a sort of jaw) that squeezes the disk when a force is applied to the fork collar (see the rendering of the disk brake in Fig. \ref{fig:rendering_velocita}a)). In Fig. \ref{fig:rendering_velocita}b), it is shown the velocity field calculated by the CFD software. The inlet (fork collar located at the bottom of the image) is the area where the actuation force is applied to. The two symmetric outlets are the areas where the proposed concept displaces toward the disk surfaces. 
The velocity field is employed to calculate the cell node displacements according to STEP 1. Results are  shown in Fig. \ref{fig:celle}. 

This solution is innovative because the actuation system is integrated within the caliper structure and it is possible, in this way, to remarkably reduce  the number of mechanical components.
In Fig. \ref{fig:rendering_velocita}a), the rendering of the disk and of the liquid structure caliper is shown.

\section{Open questions and future developments}
\label{future}
The work presented in this paper constitutes a pioneering and unconventional use of CFD codes for the design of a new family of metamaterials. It has the merit of providing a general theoretical framework for the new concept of liquid structures, but before putting them into practice in some concrete applications (e.g. in the field of MEMS and 3D printing) it is necessary to deepen some theoretical and implementation aspects. Due to the vastness of the topic, it is not possible to go into every single facet of the problem. However, it is necessary to identify some directions of investigation that may be useful for the studies that will follow. Therefore, in this section, some open questions and problems are presented.

\begin{itemize}
  \item \textit{Extension to the 3D case} 
  
As far as star cells are concerned, the extension to the 3D case is not so trivial. The cell should have 8 nodes (the vertices of the cube that inscribes it) where 8 pyramids should point at, each of them composed of 3 links. It remains to be identified, however, the internal topology able to introduce the \#DoF necessary to provide the required mobility (see Fig. \ref{fig:3Dcell}). 

\begin{figure}[h!]
\centering
    \includegraphics[width=0.6\textwidth]{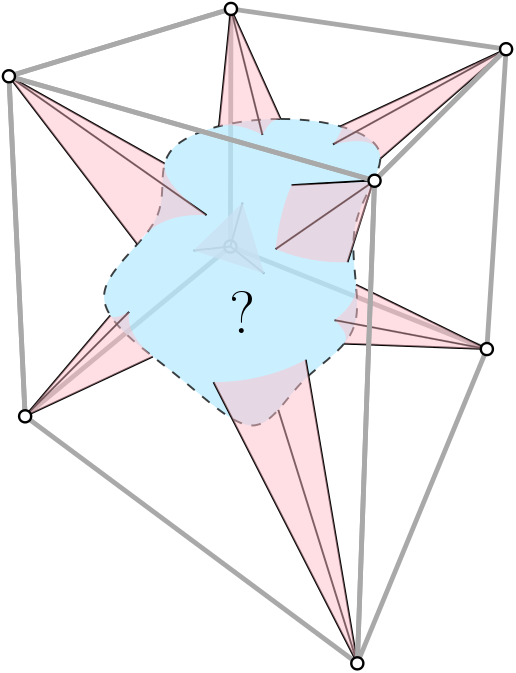}
    \caption{Sketch of a star cell in 3D with undetermined internal topology.}
    \label{fig:3Dcell}
\end{figure}

The task is even more difficult if one wants to obtain a cell with only 4 DoF. In any case, our intuition is that it is necessary to sacrifice the topological symmetry that the 2D star cells possess.

  \item \textit{Study of the kinematic properties of different cell topologies} 
  
As shown in the previous sections, there are many types of cell that can be used (as examples we introduced the synthesis of star and diamond cells). It would also be interesting to use different typologies for the same structure, choosing locally the more appropriate one according to the displacements that are imposed to the nodes of the cell.

As far as the star cell is concerned, we notice that the mobility is guaranteed by a RRRR linkage (a 1 DoF four-bar linkage). We could have the same mobility (1 DoF) by exchanging a revolute joint for a prismatic one. For example, in Fig. \ref{fig:diadi}, the mechanism is a RPRR. On a theoretical basis, all R-P permutations are allowed, although very likely not all of them have the same kinematic efficiency.
  
  \begin{figure}[h!]
\centering
    \includegraphics[width=0.7\textwidth]{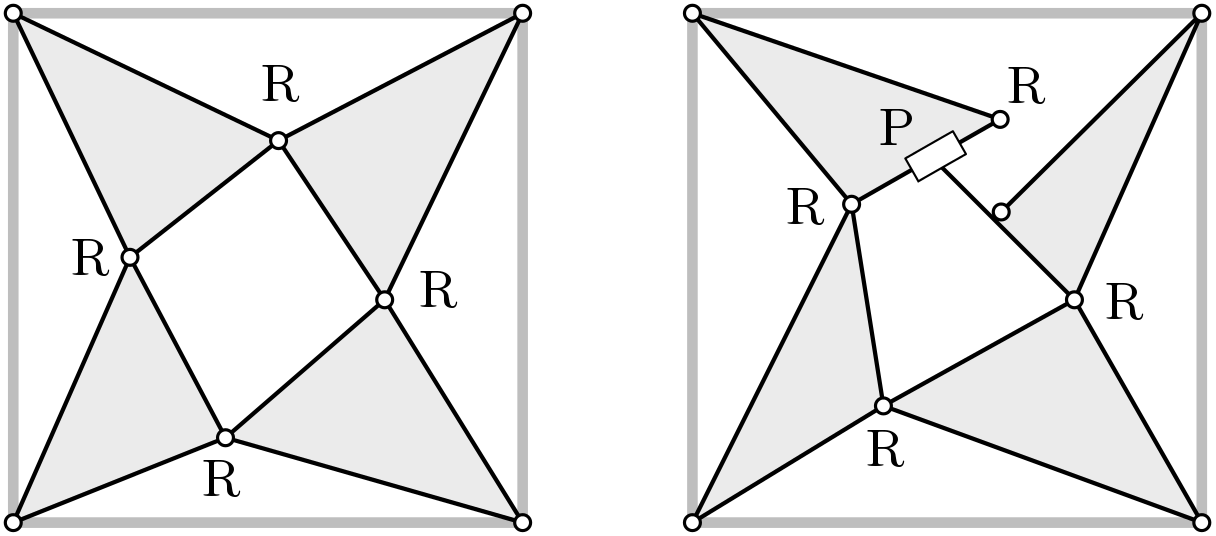}
    \caption{Two examples of star cells with different inner joints (revolute (R) and prismatic (P)).}
    \label{fig:diadi}
\end{figure}

  \item \textit{Energy required to allow the soft mode}
  
Since liquid structures are \textit{de facto} bistable mechanisms, it is necessary to calculate the value of the maximum elastic energy stored by the system during the transition (soft mode). If this energy is too high, motion difficulties may emerge. To achieve the goal, once the kinematic synthesis is completed, it is necessary to implement a static equilibrium model that takes into account the elasticity of each single link.

It is intuitive that the larger the movements at the nodes of the cells, the more energy will be stored by the mechanism during the transition. This statement suggests that alternative strategies can be investigated to mitigate this problem or to achieve overall compromises or optimizations.

 \item \textit{Alternatives to fluid flow equivalence for the calculation of node displacements}
 
In STEP 1 (Subsection \ref{STEP_1}), the nodes displacements have been calculated by exploiting the fluid flow equivalence, namely solving numerically a Navier-Stokes steady-state partial differential equation problem. This method ensures a gradual and continuous variation of the nodes displacements along the corridor area that connects the inlet nodes to the outlet ones. However, alternative strategies can be pursued. Promising directions of investigation could be topological optimizations through genetic algorithms \cite{Li20192405,Cuevas20192963,Christiansen2019}, heuristic algorithms or path-planning techniques \cite{Gallina2000237}.

\item \textit{Singular
configurations}

In the theory of mechanisms, the problem of singular configurations is well known \cite{Lai1985} \cite{Buckens1965301}. It occurs when the determinant of the associated Jacobian is null. 

Singular configurations must be avoided both because they make very difficult to operate the mechanism and because they introduce an indeterminateness in the calculation of the inverse kinematic solution.
In liquid structures this is a very serious problem, since the occurrence of a singular configuration in a cell can affect the  functionality of the whole structure.

It is not so easy to deal with the problem in a generalized perspective. In fact, singular configurations depend on the generalized coordinates of the whole problem. Each individual cell cannot be analyzed separately, but must be considered within the context of the whole mechanism. Moreover, the issues pertinent to the intrinsic elasticity of links complicate the model to be used for the analysis.

\item \textit{How to prevent the cell links from crossing each other?}

The synthesis of the cell, although correct from a kinematic point of view (the imposed constraints are fulfilled), could lead to unacceptable solutions from the construction point of view. Two problems may occur: 1) For a star cell, some links of the central 4-bar linkage could cross each other; this is an issue because the mechanism  must be realized on two different layers to prevent the links to interfere each other; 2) some links could come out the square area defined by the four internal nodes. The latter is not desirable because the cell could intersect with the adjacent ones.

\item \textit{Is it possible to have more than just a single soft mode on the same liquid structure?}

The proposed theory considers a single corridor, with a unique inlet and unique outlet. But the same liquid structure could have more inlets and more outlets. It is not trivial to determine whether criss-crossing corridors can be made.   

\item \textit{Is there a strategy to place the internal nodes in a more efficient way?}

In the provided examples, nodes are arranged along a regular grid. However, they could be placed in to make the liquid structure more efficient. For example, the nodes could be aligned along flow lines. 

\end{itemize}

\section{Conclusions}

%This paper presented the synthesis of a mechanism that constitutes a metamaterial (liquid structure) able to show soft modes. To achieve this goal, the classical synthesis of mechanisms has been used in combination with CFD analysis methods. The aim of the work was twofold: on the one hand, a solid baseline framework that can be extended by future contributions has been defined; on the other hand, open questions and future research directions that deserve to be investigated have been clearly presented for the benefit of the scientific community.  

In this paper, we presented the new concept of liquid structure, inspired by the kinematics of constant flow of an incompressible fluid within a limited domain. A liquid structure is composed of several bistable assemblies of pseudo-rigid links and joints called cells that are able to show soft modes and transmit a mechanical input between two distant portions of the domain. To achieve this goal, the classical synthesis of mechanisms has been used in combination with CFD analysis methods. The aim of the work was twofold: on the one hand, a solid baseline framework 
in which the kinematics synthesis is grounded on a two-step process has been defined; on the other hand, open questions related to theoretical and manufacturing aspects, and future research directions that deserve to be investigated have been clearly presented for the benefit of the scientific community. 

Our work defines the guidelines for a research program aiming at optimizing and realizing effective liquid structures for multi-scale applications in mechanical engineering.

\vspace{5mm}

\noindent
\textbf{Acknowledgements.}
This work has been partially supported by Italian MIUR under PRIN grant “SEDUCE” no. 2017TWRCNB and by University of Trieste under grant FRA 2018. The LAMA F.V.G. project (http://lamafvg.it/) is also gratefully acknowledged.

%\label{}

\appendix

\section{Matrix equation terms related to star and diamond cell synthesis}
\label{matrices}
As far as the star cell is concerned, the system matrix, the unknowns vector and the constant term are
\begin{multline}
\\ \boldsymbol{A}_s=\begin{bmatrix}
 -1 &1  &0  &0  &0  &0  &0  &0 \\ 
 0  &  0&  -1&  1&  0&  0&  0&0 \\ 
 0&  0&  0&  0&  1&  -1&  0&0 \\ 
 0&  0&  0&  0&  0&  0&  1&-1 \\ 
 -e^{\Delta\;\beta_1}&  e^{\Delta\;\beta_2}&  0&  0&  0&  0&  0& 0\\ 
 0&  0&  -e^{\Delta\;\beta_2}&  e^{\Delta\;\beta_3}&  0&  0&  0& 0\\ 
 0&  0&  0&  0&  -e^{\Delta\;\beta_3}&  e^{\Delta\;\beta_4}&  0& 0\\ 
 0&  0&  0&  0&  0&  0&  -e^{\Delta\;\beta_4}& e^{\Delta\;\beta_1} 
\end{bmatrix}, \\
\boldsymbol{z}_s=\begin{Bmatrix}
z_1\\ 
z_2\\ 
z_3\\ 
z_4\\ 
z_5\\ 
z_6\\ 
z_7\\ 
z_8
\end{Bmatrix},\:\:\textrm{and}\:\: 
\boldsymbol{u}_s=
\begin{Bmatrix}
h\\ 
u\\ 
h\\ 
u\\ 
h+\delta_{in\;2}-\delta_{in\;1}\\ 
u+\delta_{in\;3}-\delta_{in\;2}\\ 
-h+\delta_{in\;4}-\delta_{in\;3}\\ 
-u+\delta_{in\;1}-\delta_{in\;4}
\end{Bmatrix},\  \textrm{respectively}.
\\
\end{multline}

For the diamond cell the same quantities are
\begin{multline}
\\ \boldsymbol{A}_d=\begin{bmatrix}
 1 &0  &0  &1  &0  &0 \\ 
 0 &1  &0  &-1 &1  &0 \\ 
 0 &0  &-1 &0  &1  &-1 \\ 
 -e^{\Delta\;\beta_1}&  0&  0&  e^{\Delta\;\beta_4}& 0&  0& \\ 
 0&  e^{\Delta\;\beta_1}&  0& -e^{\Delta\;\beta_4}&  e^{\Delta\;\beta_5}&  0\\ 
 0&  0&  -e^{\Delta\;\beta_1}&  0&  e^{\Delta\;\beta_5}&  -e^{\Delta\;\beta_6} 
\end{bmatrix}, \\
\boldsymbol{z}_d=\begin{Bmatrix}
z_1\\ 
z_2\\ 
z_3\\ 
z_4\\ 
z_5\\ 
z_6
\end{Bmatrix},\:\: \textrm{and} \:\:
\boldsymbol{u}_d=
\begin{Bmatrix}
h\\ 
u\\ 
h\\ 
h-\delta_{in\;1}+\delta_{in\;2}\\ 
u-\delta_{in\;2}+\delta_{in\;3}\\ 
-h-\delta_{in\;3}+\delta_{in\;4}\\ 
\end{Bmatrix}.
\\
\end{multline}

\section{Bistability of a prototype cell}

The cell sketched in Fig. \ref{fig:bistable} is designed to achieve $\delta_1=2 \delta_2=0.4$. Coordinates of its points are $Q_1=(0,1)$, $Q_2=(1,1)$, $Q_3=(0.65,0)$, $Q_4=(0.2,0)$, $Q_5=(0.1,0.5)$ and $Q_6=(0.55,0.75)$.
Lengths of rigid bars are $a=\overline{Q_4 Q_5}=\overline{Q_1 Q_5}$, $b=\overline{Q_5 Q_6}$ and $c=\overline{Q_3 Q_6}$, and their values can be easily inferred from the coordinates just introduced. 
Representative angles of the initial configuration $C_1$ take the values $\alpha=\arctan(5)$, $\alpha_b=\arctan(5/9)$, $\alpha_c=\arctan(7.5)$ while $\beta$ and $\gamma$, on the right of the figure, are connected to $\theta$ through the set of equations
\begin{gather} \label{catenasyst}
-a \cos(\alpha+\theta)+b \cos(\alpha_b+\beta)+ c \cos (\alpha_c+\gamma)=0.45, \\
a \sin(\alpha+\theta)+b \sin(\alpha_b+\beta)- c \sin (\alpha_c+\gamma)=0. 
\end{gather}
Shortenings of springs can be written as $\eta_1(\theta)=a-l_{15}(\theta)$ and  $\eta_2(\theta)=d-l_{26}(\theta)$, where
\begin{equation}
l_{15}(\theta)=\frac{1}{\sin(\alpha+\theta)}-a,
\ \ \ 
l_{26}(\theta)= \frac{1- a \sin(\alpha+\theta)}{\sin(\alpha_b+\beta(\theta))}- b.
\end{equation}
It may be finally interesting to note that the displacements of the nodes $Q_1$, $Q_2$ along the axis of symmetry, namely $s_1(\theta)$ and $s_2(\theta)$, are 
\begin{gather}
 s_1(\theta)=0.2-  \frac{1}{\tan(\alpha+\theta)}, \\
s_2(\theta)= l_{26}(\theta) \cos(\alpha_b+\beta(\theta))-0.35-c \cos(\alpha_c+\gamma(\theta)).
\end{gather}


%%       \citet{<label>} ==> Jones et al. [21]
%%       \citep{<label>} ==> [21]
%%

\bibliographystyle{elsarticle-num-names} 
\bibliography{main_v2.bib}


%\begin{thebibliography}{00}



%\end{thebibliography}

\end{document}